\documentclass[12pt]{article}
\usepackage{amssymb}
\usepackage{graphicx}
\usepackage{amsmath}
\usepackage[all]{xy}
\usepackage{color}

\def\eeq{\end{eqnarray}}
\def\hsp{,\hspace{.7cm}}

\newcommand{\nn}{\nonumber}

\def\D{\mathcal{D}}

\def\de{\partial}

\def\=:{=\hspace{-.7em}\raisebox{1.1ex}{.}\hspace{.1em}\raisebox{-0.2ex}{.} }
\newcommand{\NF}{N_{\rm F}}
\newcommand{\NC}{N_{\rm C}}
\newcommand{\tNC}{\tilde{N}_{\rm C}}

\newcommand {\beq}{\begin{eqnarray}}
\newcommand {\eeqq}{\end{eqnarray}}

\def\nn{\nonumber\\}

\newcommand {\lam}{\lambda}

\newcommand {\Tr}{{\rm Tr}\,}

\newcommand {\Nf}{N_{\rm F}}
\newcommand {\Nc}{N_{\rm C}}

\def\mnchanged#1{#1}
\def\kochanged#1{#1}
\def\nychanged#1{#1}

\makeatletter
\@addtoreset{equation}{section}

\renewcommand{\thefootnote}{\fnsymbol{footnote}}
\renewcommand{\hat}{\widehat}
\makeatother

\def\changed#1{{#1}}

\begin{document}


\thispagestyle{empty}
\begin{flushright}
TIT/HEP-571, IFUP-TH/2007-09, ULB-TH/07-16, UT-Komaba/07-4\\
April, 2007 \\
\end{flushright}
\vspace{3mm}

\begin{center}
{\Large \bf
On the moduli space of semilocal strings and lumps}
\\[12mm]
\vspace{5mm}

\normalsize
{ \bf
Minoru~Eto}$^{1,2,3}$
,
{ \bf Jarah~Evslin}$^{3,4}$
,
{ \bf
Kenichi~Konishi}$^{2,3}$
,
{ \bf
Giacomo~Marmorini}$^{2,5}$
,
{ \bf
Muneto~Nitta}$^6$
,
{ \bf
Keisuke~Ohashi}$^7$
,
{ \bf
Walter~Vinci}$^{2,3}$
,
{ \bf
Naoto~Yokoi}$^{1}$
\footnote{e-mail: minoru(at)df.unipi.it, jevslin(at)ulb.ac.be, konishi(at)df.unipi.it, g.marmorini(at)sns.it,
nitta(at)phys-h.keio.ac.jp, walter.vinci(at)pi.infn.it, nyokoi(at)hep1.c.u-tokyo.ac.jp}

\vskip 1.5em

{\small
$^1$ {\it
Institute of Physics, The University of Tokyo,
Komaba 3-8-1, Meguro-ku
Tokyo 153-8902, Japan
}
\\
$^2$ {\it INFN, Sezione di Pisa,
Largo Pontecorvo, 3, Ed. C, 56127 Pisa, Italy
}
\\
$^3$ {\it Department of Physics, University of Pisa
Largo Pontecorvo, 3, Ed. C, 56127 Pisa, Italy
}
\\
$^4$ {\it
Physique Th\'eorique et Math\'ematique,
Universit\'e Libre de Bruxelles \\\& International Solvay
Institutes, ULB C.P. 231, B1050 Bruxelles, Belgium}
\\
$^5$ {\it Scuola Normale Superiore,
Piazza dei Cavalieri, 7, 56126
Pisa, Italy
}
\\
$^6$ {\it
Department of Physics, Keio University, Hiyoshi,
Yokohama, Kanagawa 223-8521, Japan
}
\\
$^7$ {\it Department of Physics, Tokyo Institute of
Technology,
Tokyo 152-8551, Japan
}
}

\vspace{15mm}
%
%
{\bf Abstract}\\[5mm]
{\parbox{13cm}{\hspace{5mm}

We study BPS non-abelian semilocal vortices in $U(\NC)$ gauge theory with $\NF$ flavors, $\NF>\NC$, in the Higgs phase. The
moduli space for arbitrary winding number is described using the moduli matrix formalism. We find a relation between the moduli
spaces of the semilocal vortices in a Seiberg-like dual pairs of theories, $U(\NC)$ and $U({\NF-\NC})$. They are two
alternative regularizations of a ``parent'' non-Hausdorff space, which tend to the same moduli space of sigma model lumps in the
infinite gauge coupling limits. We examine the normalizability of the zero-modes and find the somewhat surprising
phenomenon that the number of normalizable zero-modes, dynamical fields in the effective action, depends on the
point of the moduli space we are considering. We find, in the lump limit, an effective action on the vortex worldsheet, which we
compare to that found by Shifman and Yung.


}}
\end{center}
\vfill
\newpage
\setcounter{page}{1}
\setcounter{footnote}{0}
\renewcommand{\thefootnote}{\arabic{footnote}}

\section{Introduction and discussion }\label{Intro}

Solitons in classical and quantum field theories (and string theory) have always attracted interest due to their applications to
numerous branches of physics. Vortices in particular play a crucial role in many different contexts, from condensed matter to
high energy physics and cosmology.

While abelian vortices \cite{Abrikosov:1956sx,Nielsen:1973cs} have been widely studied in the literature, non-abelian vortices
were introduced quite recently \cite{Auzzi:2003fs,Hanany:2003hp}. These configurations are characterized by non-abelian
zero-modes (moduli) related to their orientation in the internal color-flavor space. Since the introduction of the moduli matrix
formalism (for a review see \cite{Eto:2006pg}), the study of non-abelian vortices has acquired new emphasis: the characterization
and the analysis of their moduli spaces in terms of moduli matrix parameters \cite{Eto:2005yh,Eto:2006cx} has contributed to new
insights into non-abelian electric-magnetic duality \cite{Eto:2006dx} and the issue of reconnection of non-abelian cosmic strings
\cite{Eto:2006db}. A review of solitons containing non-abelian vortices can be found in
\cite{Eto:2006pg,Tong:2005un,Konishi:2007dn,Shifman:2007ce}.  Here we extend the investigation of this class of solitons to the
semilocal case.

The term {\it semilocal vortex} was invented for string-like objects in abelian Higgs models with more than one Higgs field
\cite{Vachaspati:1991dz}, where a global (flavor) symmetry group is present in addition to the local (gauge) symmetry group. In
the non-abelian context of interest in this note, if the gauge group is $U(\NC)$ and there are $\NF$ flavors of fundamental
matter, {\it semilocal} will refer to the case $\NF >\NC$. Semilocal abelian vortices are known to exhibit peculiar properties,
which are very different from those of the usual Abrikosov-Nielsen-Olesen (ANO) vortices \cite{Abrikosov:1956sx,Nielsen:1973cs}.
As usual their total magnetic flux is quantized in terms of their topological charge, however the magnetic field of a semilocal
vortex does not decay exponentially in the radial direction, instead it falls off according to a power law. Moreover the
transverse size of the flux tube is not fixed but becomes a modulus. This feature gives rise to questions about the stability of
these objects; in \cite{Hindmarsh:1991jq} (see also \cite{Achucarro:1999it}) it is argued that they are stable if the quartic
coupling in the potential is less than or equal to the critical (BPS) value, {\it i.e.}, if the mass of the scalar is less than
or equal to the mass of the photon.

Semilocal vortices interpolate between ANO vortices and sigma model lumps \cite{Hindmarsh:1991jq,Schroers:1996zy}, to which they
reduce in two different limits. It is possible to study their dynamics \cite{Leese:1992fn} in the moduli space approximation \mnchanged{\cite{Manton:1981mp}} and
also in the lump limit \cite{Ward:1985ij}. It turns out that, in general, the fluctuations of some zero-modes corresponding to
the global size of the configuration actually cost infinite energy. These zero-modes have to be fixed to make moduli space
dynamics meaningful.

It is natural to ask what emerges for semilocal non-abelian vortices.
\mnchanged{
The number of zero modes, namely the dimension of the moduli space,
of winding number $k$ was calculated to be $2kN_{\rm F}$ in \cite{Hanany:2003hp}.
}
The problem
\mnchanged{of (non-)normalizability of these zero modes, or the construction of the effective theory,}
has been considered in detail for ${\cal N}=2$ supersymmetric $U(2)$ gauge theories with $\NF=3,4$ in \cite{Shifman:2006kd} by
Shifman and Yung. They found BPS solutions for single non-abelian semilocal vortices and then they used symmetry arguments to
develop an effective theory on the vortex worldsheet. They noted that single semilocal vortices have only non-normalizable
zero-modes
\mnchanged{except for the position modulus}:
not only the size modulus, but also the orientational moduli undergo this pathology, which is somewhat more surprising.
This behavior is manifest because the effective worldsheet theory is a two dimensional sigma model, whose target space has a
divergent metric unless an infrared regulator is provided. In this respect the geodesic motion on the moduli space seems
essentially frozen, in contrast with the local case.

The aim of this paper is to generalize the moduli matrix approach \cite{Eto:2006pg} to semilocal vortices
(a first application of this method to semilocal strings is found in \cite{Isozumi:2004vg}).
Our considerations \mnchanged{here} apply to vortex configurations of generic winding number $k$ in
\nychanged{any $U(\NC)$ gauge theory with $\NF$ flavors and the critical quartic coupling, in the case of $\NF>\NC$}.
We analyze their moduli space both at the kinematical and at the dynamical level. First we provide an unambiguous smooth
parametrization of the moduli space, provided by the moduli matrix, and study its \mnchanged{topological} structure \mnchanged{(without the metric)}.
\mnchanged{The moduli metric is defined only for normalizable moduli
which parameterize a subspace inside the whole moduli space
of dimension $2kN_{\rm F}$.}
We use supersymmetry
to derive an effective action for the system of $k$ vortices and show that, even though a single semilocal vortex always has only
non-normalizable moduli \cite{Shifman:2006kd}, higher winding configurations admit normalizable moduli, which roughly correspond to relative sizes and
orientations in the internal space, \mnchanged{as first noted in \cite{Eto:2006db}}. This means that, upon fixing the non-normalizable global moduli, the analysis of the geodesic
motion on the moduli space becomes meaningful. Along the way we discover an interesting relationship between the vortex moduli
spaces of a Seiberg-like dual pair of theories at fixed winding number $k$: they both descend, by means of two alternative regularizations, from the same ``parent space'', which is a non-Hausdorff space defined in terms of a certain holomorphic quotient. As a result they are guaranteed to be birationally equivalent and in fact they are related by a geometric transition. Moreover, in the limit
of infinite gauge coupling we see that they reduce to the same moduli space of $k$ {Grassmannian} sigma model lumps, as is expected on general grounds. We also find the normalizable moduli enhancement on special submanifolds: the number of normalizable moduli can change depending on the point of moduli space we are dealing with.
For instance, the orientational moduli ${\bf C}P^{\NC-1}$
of a $k=1$ vortex are non-normalizable unless the size modulus vanishes.
However they become normalizable in the limit of vanishing size modulus
where the semilocal vortex shrinks to a local vortex
(with physically non-zero size).

The paper is organized as follows. In Section~\ref{setup} we define the model, write down BPS vortex equations and review the moduli matrix formalism for analyzing the set of solutions. In Section~\ref{msq} we discuss how the various sectors of the vortex moduli space are described by holomorphic or symplectic quotients with specific properties; we also describe how the $k$-th topological sector and that of a Seiberg-like dual theory are related. This pair of dual spaces becomes the same Grassmannian lump moduli space at infinite gauge coupling, as is explained in Section~\ref{lumplim}; there, we also find an interesting extension of the holomorphic rational map approach to Grassmannian lumps. Section~\ref{example} is devoted to the presentation of explicit examples. In Section~\ref{effact} we \mnchanged{examine (non-)normalizability} of zero-modes and \mnchanged{obtain}
the worldsheet effective action; some results are compared with those previously obtained by Shifman and Yung. Finally in Section~\ref{conclusion} a summary and conclusions are given.
\nychanged{In Appendix~\ref{wpsnw}, some geometrical properties of weighted
projective spaces with both positive and negative weights are
briefly summarized. The detailed analysis for the moduli spaces of
$k=2$ composite semilocal vortices is given in Appendix~\ref{compsemvort},
and the moduli space for lumps is obtained in terms of
the moduli matrix in Appendix~\ref{Ohashi}.}

\section{The $U(\NC)$ model and the moduli matrix} \label{setup}

We shall be interested in a $U(\NC)$ gauge theory with $\NF$ flavors of fundamental scalars, which we collect in an $\NC$ by $\NF$
matrix $H$. We restrict our attention to the case $\NF>\NC$, where semilocal vortices are admitted. The Lagrangian of this gauge theory is
\beq
{\cal L} = \Tr \left[ - \frac{1}{2g^2} F_{\mu\nu}\, F^{\mu\nu} - \D_\mu \,H \;\D^\mu H^\dagger - \lambda \left( \xi \,{\bf 1}_{N_{\rm C}} - H\,H^\dagger\right)^2\right] \label{lagrangian}
\eeqq
where we have defined the field strength and covariant derivative
\beq
F_{\mu\nu} = \partial_\mu W_\nu - \partial_\nu W_\mu + i \left[W_\mu,W_\nu\right]\hsp \D_\mu H = \left(\partial_\mu + i\, W_\mu\right)\, H
\eeq
in terms of the connection $W_\mu$. Here $g$ is the $U(\NC)$ gauge coupling and $\lambda$ is the scalar quartic coupling
constant. Throughout this paper we shall take the critical (BPS) value $\lambda=g^2/4$, which assures that our Lagrangian is the
bosonic sector of a supersymmetric model.\footnote{
\mnchanged{
The Lagrangian (\ref{lagrangian}) possesses ${\cal N}=2$ supersymmetry
if we suitably add two adjoint scalar fields, another $\NC$ by $\NF$ matrix of (anti-)fundamental scalars and}
\nychanged{all the superpartners}.}
In the supersymmetric context $\xi$ is interpreted as the Fayet-Iliopoulos parameter. We
will set $\xi>0$, so as to have stable vortex configurations. The Lagrangian (\ref{lagrangian}) is invariant under a global
$SU(\NF)$ flavor symmetry which acts on $H$ from the right. The vacuum equation $H\, H^\dagger = \xi\,{\bf 1}_{\NC}$ implies that
$H$ has maximal rank and so, after quotienting out the gauge equivalent configurations, $H$ defines a Grassmannian manifold.
Therefore the model has a continuous Higgs branch
\beq {\cal V}_{\rm Higgs} = Gr_{\NF,\NC}
 \mnchanged{\simeq {SU(\NF) \over SU(\NC) \times SU(\NF-\NC)\times U(1)}}
 \label{Higgs-vac}
\eeq
and no Coulomb vacuum.\footnote{
\mnchanged{
In the context of ${\cal N}=2$ supersymmetry the Higgs branch is the cotangent bundle over the Grassmannian manifold, $T^*Gr_{\NF,\NC}$ \cite{Lindstrom:1983rt}
obtained as a hyper-K\"ahler quotient \cite{Hitchin:1986ea}.
}
}
The gauge symmetry is completely broken while an exact global $SU(\NC)_{\rm C+F}$ color-flavor symmetry
remains unbroken.
\mnchanged{
We define a dual theory by the same Lagrangian
(\ref{lagrangian}) with different gauge group
\nychanged{$U(\tilde \NC)$} ($\tilde \NC \equiv \NF-\NC$)
and the same number of flavors.
From the last expression of (\ref{Higgs-vac})
the Higgs branch of the dual theory is obviously identical to
that of the original theory.
We \nychanged{refer to this duality as ``Seiberg-like dual", or
simply ``Seiberg dual''}.
 In the Hanany-Witten type D-brane realization \cite{Hanany:1996ie}
this duality can be understood as
exchange of two NS5-branes,
while the original Seiberg duality in ${\cal N}=1$ theory
can be understood by this procedure with
one NS5-brane rotated.
}

This system has non-abelian vortex solutions which satisfy the BPS equations
\beq
\left (\D_1+i\D_2\right) \, H = 0,\quad F_{12} + \frac{g^2}{2} \left( \xi \,{\bf 1}_{N_{\rm C}} - H\, H^\dagger\right) =0.
\label{eq:bps_equation}
\eeqq
Actually these equations possess a continuous moduli space of solutions. Since every solution is characterized by a topological
(winding) number valued in $\pi_1(U(\NC))=\mathbb{Z}$, the moduli space is divided into topological sectors. Upon introducing a
complex parametrization of the plane, $z=x_1+ix_2$, coordinates on the moduli space are conveniently collected in a unique
mathematical object $H_0(z)$, a holomorphic $\NC\times\NF$ matrix called the moduli matrix which is defined by \cite{Isozumi:2004vg}
\beq
H = S^{-1}(z,\bar z) \, H_0(z),\quad
W_1 + i\,W_2 = - 2\,i\,S^{-1}(z,\bar z) \, \bar\partial_z S(z,\bar z)
\label{bpssol}
\eeqq
where $S(z,\bar z)$ is an $\NC\times \NC$ {\it invertible} matrix.
The elements of $H_0(z)$ are polynomials in $z$ whose coefficients
are good coordinates (in the sense of \cite{Eto:2006db}) on the moduli space. A configuration has winding number $k$ if
\beq \label{windk}
\det H_0H_0^{\dagger}={\cal O}(|z|^{2k})
\eeq
for large $z$. From the definition~(\ref{bpssol}) one sees that $H_0(z)$ and $S(z,\bar{z})$ are only determined up to the
so-called $V$-equivalence given by
\beq H_0(z) \to V(z) \, H_0(z), \quad S^{-1}(z,\bar{z}) \to S^{-1}(z,\bar{z})\, V(z)^{-1}\label{Vequivalence}
\eeqq
where $V(z)$ is a $GL(\NC,{\bf C})$ matrix \nychanged{whose} \kochanged{elements
are polynomials in $z$}.
Eq.~(\ref{windk}) implies that if one fixes the winding number $k$, then
$\det V$ must be \kochanged{constant}.

The first BPS equation is automatically solved by the ansatz (\ref{bpssol}), while the second can be rewritten as \cite{Isozumi:2004vg}
\beq \partial_z (\Omega^{-1}\bar{\partial}_z \Omega)
= \frac{g^2}{4}\left(\xi {\bf 1}_{N_{\rm C}} -\Omega^{-1}H_0H_0^\dagger\right), \label{mastereq}
\eeq
where $\Omega\equiv S(z,\bar{z})S^\dagger (z, \bar{z})$. We will refer to Eq.~(\ref{mastereq}) as the master equation, and assume
that it has a unique solution \kochanged{with a boundary condition
$\Omega\rightarrow \xi^{-1}H_0H_0^\dagger$}. This assumption has only been proven in the abelian case and for vortices on compact Riemann
surfaces; however, there are arguments that it extends to general vortices on ${\bf C}$ (see \cite {Eto:2006pg} for a more
detailed discussion).

Following \cite{Eto:2006pg}, it is possible to organize the moduli into a K\"ahler quotient \mnchanged{\cite{Hitchin:1986ea}}. First, let us write the moduli
matrix as
\beq \label{decomp}
H_0(z)=\left({\bf D}(z),{\bf Q}(z)\right )
\eeq
where ${\bf D}(z)$ is \mnchanged{an} $\Nc \times \Nc$ matrix and ${\bf Q}(z)$ is a $\NC \times \tilde{N}_{\rm C}$ matrix ($\tilde{N}_{\rm
C}\equiv \NF-\NC$). Defining
\beq P(z)\equiv \det {\bf D}(z), \eeqq
 we set $\deg P(z)=k$, whereas all other minor determinants
of $H_0(z)$ have degree at most $k-1$: this guarantees that Eq.~(\ref{windk}) is satisfied and the winding number is equal to
$k$. Moreover, using the Pl\"ucker relations (see, for instance, Eq.~(\ref{pluecker})), one finds that only a subset of the minor determinants are independent, namely
$P(z)$ and the determinants of the minor matrices obtained by substituting the $r$-th column of ${\bf D}(z)$, $r=1,\ldots,\NC$,
with the $A$-th column of ${\bf Q}(z)$, $A=1,\ldots,\tNC$. The matrix having \nychanged{these minor determinants
as its $(r\,A)$ elements is denoted by} ${\bf F}(z)$,
\beq {\bf F}_{r A} = \sum_{k=1}^{\NC} Q_{k A}\, ({\rm Cof} \, {\bf D})_{r k}= P(z)\, \sum_{k=1}^{\NC} Q_{k A}\, ({\bf D}^{-1})_{r k}, \qquad
{\bf F}(z) = P(z)\, {\bf D}^{-1}\,{\bf Q}(z).
\label{defF} \eeqq

Consider the equation
\beq {\bf D}(z) \vec{\phi}(z) = \vec{J}(z) P(z) =0 \mod P(z) \label{poldiv}
\eeq
for $\vec\phi$ modulo $P(z)$, where the components of $\vec\phi$ are polynomials at most of degree $z^{k-1}$. We find $k$ linearly independent such vectors $\vec{\phi}_i(z)$, $i=1,\ldots,k$, each of which is a solution
with a suitable $\vec{J}_i(z)$. In matrix form,
\beq {\bf D}(z) {\bf\Phi}(z) = {\bf J}(z) P(z) =0 \mod P(z)
\label{eigenmatrix1}
\eeq
where ${\bf\Phi}(z)$ and ${\bf J}(z)$ are $\NC \times k$ matrices made of the $\{\vec{\phi}_i\}$ and the $\{\vec{J}_i\}$
respectively. We are naturally free to choose a basis in the linear space of solutions to Eq.~(\ref{poldiv}) using the
equivalence relation
\beq {\bf \Phi(z)} \sim {\bf \Phi(z)} \, {\cal V}^{-1}, \quad {\cal V} \in GL(k,{\bf C}). \label{shuffle}
\eeq
Since ${\bf \Phi(z)}$ has the maximal rank, the $\{\vec{\phi}_i\}$ being linearly independent, the $GL(k,{\bf C})$ action is free\footnote{A group $G$ is said to act freely on a space $M$, if for any point $x \in M$, $g \, x = x$
($g \in G$)
implies $g= {\bf 1}$. }
\beq {\bf \Phi(z)} ={\bf \Phi(z)} {\cal V}^{-1} \Rightarrow {\cal V}^{-1}={\bf 1}_k. \label{phifree}
\eeq

Now consider the product $z \, \vec{\phi}_i(z)$, whose degree is in general less than or equal to $k$. The polynomial division by
$P(z)$ leads to a constant quotient and a remainder that must be a linear combination of $\{\vec{\phi}_i\}$, as it must
 satisfy Eq.~(\ref{poldiv}). In matrix form,
\kochanged{these are summarized as
\beq z \, {\bf\Phi}(z) = {\bf\Phi}(z) {\bf Z} + {\bf \Psi} P(z).\label{poldiv2}
\eeq
By multiplying ${\bf D}(z)/P(z)$ from \nychanged{the left} and
using Eq.(\ref{eigenmatrix1}), we also find,
\begin{eqnarray}
 z {\bf J}(z)={\bf J}(z){\bf Z}+{\bf D}(z){\bf \Psi}.\label{halfADHM}
\end{eqnarray}
}
This defines uniquely the constant matrices ${\bf Z}$ and ${\bf \Psi}$,
of \mnchanged{sizes} $k\times k$ and $\Nc \times k$ respectively.
They enjoy an equivalence relation due to (\ref{shuffle})
\beq \left({\bf Z},\, {\bf \Psi}\right) \sim \left({\cal V}\,{\bf Z}\,{\cal V}^{-1}, \, {\bf \Psi}\,{\cal V}^{-1} \right), \qquad {\cal V} \in GL(k,{\bf C}),
\label{eqv1}\eeq
where the $GL(k,{\bf C})$ action is free (Eq.~(\ref{phifree})).
\kochanged{Eigenvalues of ${\bf Z}$ describe $k$ positions of vortices, and
thus there is an equality $P(z)=\det {\bf D}(z)=\det(z-{\bf Z})$.
Roughly speaking, each column \nychanged{of $\Psi$} parametrizes an orientation \nychanged{in}
${\bf C}P^{N_{\rm C}-1}$ of each
corresponding vortex.}
This would be the whole story in the local case, but in
the semilocal case there are extra moduli coming from ${\bf Q}(z)$.

Using the relation
\beq \mathbf{D}(z) \mathbf{F}(z)=\mathbf{Q}(z) P(z) \label{eigenmatrix2}
\eeq
which follows from Eq.~(\ref{defF}) and the condition $\deg {\bf F}(z)_{r A} \le k-1$ one finds that the columns of ${\bf F}(z)$ are linear combinations of those of
${\bf \Phi}(z)$,
\beq {\bf F}(z)={\bf \Phi}(z) \tilde{{\bf \Psi}}, \eeqq
 where $\tilde{{\bf \Psi}}$ is a $k\times \tilde{N}_{\rm C}$
constant matrix. Comparing Eq.~(\ref{eigenmatrix1}) and Eq.~(\ref{eigenmatrix2}) we find
\beq {\bf Q}(z)={\bf J}(z) \tilde{{\bf \Psi}}. \label{qjtildepsi}
\eeq
As ${\bf \Phi}(z)$ is defined modulo equivalence relation Eq.~(\ref{shuffle}), it follows that
$\tilde{{\bf \Psi}}$ is also defined up to
\beq \tilde{{\bf \Psi}} \sim {\cal V}\, \tilde{{\bf \Psi}}, \qquad {\cal V} \in GL(k,{\bf C}). \label{eqv2} \eeqq
All the moduli can thus be collected in the set of constant matrices $\{{\bf Z},{\bf \Psi},\tilde{{\bf \Psi}}\}$ modulo the $GL(k,{\bf C})$
equivalence of Eqs.~(\ref{eqv1}) and~(\ref{eqv2}).

\section{Moduli spaces and quotients} \label{msq}
Recently the moduli space of vortex configurations has been constructed in terms of quotient spaces. This result was achieved
both in the D-brane \cite{Hanany:2003hp} and the pure field theory approach \cite{Eto:2005yh}.

The latter approach is based on the moduli matrix formalism which, as was reviewed in Section~\ref{setup}, allows one to extract
all the moduli of BPS vortex equations from a single holomorphic matrix $H_0(z)$. For configurations of winding number $k$ in a
$U(\NC)$ gauge theory with $\NF$ fundamental flavors, the moduli are conveniently collected into the triplet $({\bf Z}, {\bf
\Psi}, \tilde{\bf \Psi})$, where ${\bf Z}$ is a $k \times k$, ${\bf \Psi}$ \mnchanged{an} $\NC\times k$ and $\tilde{\bf \Psi}$ a $k \times
\tNC$ complex matrix. They are defined modulo the $GL(k,{\bf C})$ equivalence relation
\beq
\left({\bf Z}, {\bf \Psi}, \tilde{\bf \Psi}\right) \sim \left({\cal
V}{\bf Z}{\cal V}^{-1}, {\bf \Psi}{\cal V}^{-1}, {\cal
V}\tilde{\bf \Psi}\right), \qquad {\cal V} \in GL(k,{\bf C}). \label{GLaction}
\eeq
The $GL(k,{\bf C})$ action is free on the set $\{{\bf Z}, {\bf \Psi}, \tilde{\bf \Psi}\}$, in fact it is even free on the subset
$\{{\bf Z}, {\bf \Psi}\}$. This is enough to define a good K\"ahler quotient
\mnchanged{\cite{Hitchin:1986ea}} and, indeed, the $k$-vortex moduli space turns out
to be
\beq {\cal M}_{\NC,\NF;k}= \{({\bf Z}, {\bf \Psi}, \tilde{\bf \Psi}):GL(k,{\bf C}) \hbox{ free on } ({\bf Z}, {\bf \Psi})\}/GL(k,{\bf C}). \label{Kquotient}
\eeq

Let us, instead, consider the quotient
\beq
\widehat {\cal M}_{\NC,\NF;k} \equiv
\{{\bf Z}, {\bf \Psi}, \tilde{\bf \Psi}\} / GL(k,{\bf C}) \label{quotient}
\eeq
where the $GL(k,{\bf C})$ acts freely. Now, while any free action of a compact group produces a reasonable quotient, this is not
always the case for a non-compact group, like $GL(k,{\bf C})$. The corresponding quotient can indeed present some pathologies:
in particular it \nychanged{becomes} typically non-Hausdorff \cite{Witten:1993yc}. The absence of the Hausdorff property may appear to be just a
mathematical detail but it is actually crucial to the physics. As is well known, in certain kinematical regimes, the dynamics of
solitons (and vortices among them) can be described by geodesic motion on their moduli space. If this moduli space is
non-Hausdorff, two distinct points may happen to lie at zero relative distance, in such a way that a geodesic can end at, or
simply touch, both of them at once. This is physically meaningless because two different points in the moduli space correspond to
two distinguishable physical configurations.

In general, it is possible to ``regularize'' a non-Hausdorff quotient space ({\it i.e.}, to make it Hausdorff) by removing some
points ($GL(k,{\bf C})$ orbits for us). This can be done in more than one way; indeed, as intuition suggests, if two
distinct points do not have disjoint neighborhoods, one could remove either one point or the other.

Let us describe this phenomenon from another point of view. It is possible to associate to the quotient Eq.~(\ref{quotient}) a moment map $D$,
\beq D=[{\bf Z}^\dagger,{\bf Z}]+ {\bf \Psi}^\dagger {\bf \Psi} -{\bf \tilde{\Psi}}{\bf \tilde{\Psi}}^{\dagger}-r \label{momentmap}.
\eeq
Setting $D=0$, which corresponds to fixing the imaginary part of the gauge group $GL(k,{\bf C}) \mnchanged{= U(k)^{\bf C}}$, and further dividing by the real part,
which is $U(k)$, leads to the symplectic quotient
\beq \{{\bf Z}, {\bf \Psi}, \tilde{\bf \Psi} | D=0 \} /U(k). \label{symplecticquotient}
\eeq
Now, the symplectic quotient depends on the value of $r$ in Eq.~(\ref{momentmap}). In particular, its topology is
related to the sign of $r$. There are three cases: $r>0$, $r<0$ and $r=0$, which represent three possible regularizations
of the space~(\ref{quotient}), {\it i.e.}, three possible ways to obtain Hausdorff spaces. In fact, if we choose $r=0$, the point
$({\bf Z}, {\bf \Psi}, \tilde{\bf \Psi})=(0,0,0)$, which would be a fixed point of $U(k)$, will be an element of
(\ref{symplecticquotient}). This point would have to be excluded by hand.
\mnchanged{It corresponds to a small lump singularity as discussed below.}
The choice $r \neq 0
$ guarantees a non-singular space automatically.

A large class of examples of such quotients consists of {\it weighted projective spaces}
(see Appendix~\ref{wpsnw}).
Consider for instance the simple example
 $W{\bf C}P^1_{(1,-1)}$ (Appendix~\ref{wcp1-1}). This is the space
 $\{y_1,y_2\}/{\bf C}^*$ defined by the equivalence relations $(y_1,y_2)\sim (\lambda y_1, \lambda^{-1} y_2),\; \lambda\in {\bf C}^*$. After
removing the origin $(0,0)$ the remaining sick points are $(0,y_2)$ and $(y_1,0)$, which are each in every open neighborhood that
contains the other. The two possible regularizations are:
\begin{itemize}
\item[i)] $W{\bf C}P^1_{(\underline{1},-1)}=\{(y_1,y_2) | y_1\neq 0\} / {\bf C}^*$

Introducing the moment map $D=|y_1|^2-|y_2|^2-r$ with $r>0$, $W{\bf C}P^1_{(\underline{1},-1)}$ is seen to be equivalent to the symplectic
$U(1)$ quotient $\{D=0\}/U(1)$. We have introduced a notation in which the underlined coordinates are the ones which cannot
all vanish. This is because $D$ restricted to a single ${\bf C}^*$ orbit,
\beq \tilde{D}(\lambda)=|\lambda|^2 |y_1|^2- |\lambda|^{-2}|y_2|^2-r, \label{analogous}
\eeq
is a monotonic function of $|\lambda|$; for $|\lambda|\to +\infty$, $\tilde{D}$ goes to $+\infty$ and for $|\lambda|\to 0$
it goes to $-\infty$ or $-r$ if $y_2\ne 0$ or $y_2=0$, \mnchanged{respectively}. This implies that there is a unique value of $|\lambda|$ which gives
$\tilde{D}=0$, unambiguously fixing the imaginary part of \nychanged{$U(1)^{{\bf C}} = {\bf C}^*$}.

Note that the ${\bf C}^*$ action is free on the set $\{y_1\}$, as the point $y_1=0$ is excluded.

\item[ii)] $W{\bf C}P^1_{(1,\underline{-1})}=\{(y_1,y_2) | y_2\neq 0\} / {\bf C}^*$

This is, instead, equivalent to the symplectic $U(1)$ quotient obtained by setting $r<0$ in the moment map of i).

Now the ${\bf C}^*$ action is free on the set $\{y_2\}$.
\end{itemize}
Both i) and ii) turn out to be isomorphic to ${\bf C}$, but, as mentioned above, two different regularizations of a
complex quotient lead in general to different spaces (see Subsection~\ref{single} and Appendix~\ref{wpsnw}). Note that in the case
with $r=0$, the fixed point $(0,0)$ will be a solution of (\ref{analogous}) and the resulting space will be a singular conifold.
This conifold is resolved into a regular space by setting $r \neq 0$. Similar phenomena occur in the general case of
(\ref{symplecticquotient}).

Based on considerations similar to those above, we are led to claim that, for any $k$, the $k$-vortex moduli spaces of two Seiberg-like dual theories (in the sense of Section~\ref{setup}) correspond to the two different regularization of the parent space in Eq.~(\ref{quotient}); these regularized spaces appear after
a symplectic reduction as the quotients Eq.~(\ref{symplecticquotient}) with $r>0$ and $r<0$ respectively. Indeed, in the
Seiberg dual theory the representations of the moduli $\{{\bf Z},{\bf \Psi},{\bf \tilde{\Psi}}\}$ under $GL(k,{\bf C})$ are
replaced by their complex conjugates, which is formally equivalent to flipping the sign of the Fayet-Iliopoulos parameter in
Eq.~(\ref{momentmap}).


In fact, adding the condition that $GL(k,{\bf C})$ is free on the subset $\{{\bf Z},{\bf \Psi}\}$
(resp. $\{{\bf Z},{\bf \tilde{\Psi}}\}$) to Eq.~(\ref{quotient}), as imposed by the moduli matrix construction of Section \ref{setup}, turns out to be equivalent to selecting the specific regularization
corresponding to the symplectic quotient (\ref{symplecticquotient}) with $r>0$ (resp. $r<0$), that was first found in brane theory \cite{Hanany:2003hp}.
The quotient
\beq {\cal M}_{N_{\rm C},\NF;k}= \{({\bf Z}, {\bf \Psi}, \tilde{\bf \Psi}):GL(k,{\bf C}) \hbox{ free on } ({\bf Z}, {\bf \Psi})\}/GL(k,{\bf C}). \label{theorem1}
\eeq
is isomorphic to the symplectic quotient
\beq \{({\bf Z}, {\bf \Psi}, \tilde{\bf \Psi}) : D=[{\bf Z}^\dagger,{\bf Z}]+ {\bf \Psi}^\dagger {\bf \Psi} -{\bf \tilde{\Psi}}{\bf \tilde{\Psi}}^{\dagger}-r=0 \}
/U(k). \label{theorem2}
\eeq
 with $r>0$.

Obviously, an analogous result holds with the following substitutions:
\begin{enumerate}
  \item $\NC \rightarrow \tilde{\NC} \;{(= \NF-\NC)}$
  \item $GL(k,{\bf C}) \hbox{ free on } ({\bf Z}, {\bf \Psi}) \rightarrow GL(k,{\bf C}) \hbox{ free on } ({\bf Z}, {\bf \tilde{\Psi}})$
  \item $r>0 \rightarrow r<0$.
\end{enumerate}

\section{The lump limit} \label{lumplim}
Lumps are well-known objects, arising as static finite energy configurations of codimension two in non-linear sigma models (see,
for example, \cite{Manton:2004tk}). Lumps typically are partially characterized by a size modulus, which in particular implies
that the set of lump solutions is closed with respect to finite rescaling. Formally one would also include the solution with
vanishing size modulus, but a physical configuration of zero width (and an infinitely spiked energy density) makes no sense and
must be discarded. Such limiting situations are known as small lump singularities, and they actually represent singularities in
the moduli space of lumps, which is then geodesically incomplete. In contrast, semilocal vortices do not present this kind of
pathology because they have a minimum size equal to the ANO radius $1/g\sqrt\xi$.

The situation is particularly clear for $\mathbf{C}P^1$ lumps
\mnchanged{(related to our model with $\NC=1$ and $\NF=2$)}.
In fact, the set of lump configurations in the two dimensional
$\mathbf{C}P^1$ sigma model was found to be in one-to-one correspondence with the set of holomorphic rational maps of the type
\mnchanged{\cite{Polyakov:1975yp}}
\beq R(z)= \frac{p(z) }{ q(z)}
\eeq
where $p(z)$ and $q(z)$ are two polynomials with no common factors and $\deg p<\deg q$. The topological charge of the
configuration $\pi_2({\bf C}P^1)=\mathbb{Z} $ is given by $\deg q$. It is clear that, in
order to define a fixed topological sector of the lump moduli space, one must consider the space of pairs of polynomials
$E=\{(p(z),q(z))\}$ of appropriate degree, subject to the constraint that the resultant is non-vanishing,
\beq \hbox{Res}\,[p(z),q(z)] \ne 0, \quad{\textrm{or equivalently,}\quad
|p(z)|^2+|q(z)|^2\not =0\quad {}^\forall z.} \label{resultant}
\eeq This condition excludes the singular points at which $p(z)$ and
$q(z)$ share a common factor, implying that the corresponding state is not physical\footnote{These considerations can be extended
also to general ${\bf C}P^n$ lumps using holomorphic rational maps.}.

In the limit of infinite gauge coupling, $g^2\to \infty$, our model Eq.~(\ref{lagrangian})
(the Higgs coupling is also taken to infinity $\lambda =g^2/4 \to \infty$)
reduces to a sigma model whose target
space is the Higgs branch ${\cal V}_{\rm Higgs}=Gr_{\NF,\NC}$. On the other hand, semilocal vortices therein reduce, upon
compactification of the $z$ plane, to Grassmannian lumps \cite{Hanany:2003hp,Tong:2005un}, which are topologically supported by
$\pi_2(Gr_{\NF,\NC})=\mathbb{Z}$.
\kochanged{A rational map in this case is extended to holomorphic $N_{\rm
C}\times \widetilde N_{\rm C}$ matrix given by
\begin{eqnarray}
 {\bf R}(z)\equiv\frac{1}{{\bf D}(z)}{\bf Q}(z)
=\frac{{\bf F}(z)}{P(z)}
={\bf \Psi}\frac{{\bf 1}_{k}}{z-{\bf Z}}{\bf \widetilde \Psi},
\label{rationalmap}
\end{eqnarray}
which is invariant under the $V$-transformation (\ref{Vequivalence})
and gives a holomorphic map from $S^2$ to $Gr_{\NF,\NC}$.}
Since $Gr_{\NF,\NC}=Gr_{\NF,\Nf-\NC}$, the Seiberg dual theory of Eq.~(\ref{lagrangian}) is the
same Grassmannian sigma model in the dual infinite gauge coupling limit, $\tilde{g}^2\to \infty$; moreover, its semilocal vortex
configuration tends to the same lump solutions.
\kochanged{
Actually, the extended rational map in the last form in
Eq.(\ref{rationalmap}), which is obtained by using Eq.(\ref{qjtildepsi}) and Eq.(\ref{halfADHM}), is manifestly invariant under the Seiberg-like duality.}
Of course the two dual limits cannot physically co-exist: here we are interested
in the mathematical correspondences among moduli spaces of topological string-like objects of two dual theories in the various
limits.\footnote{
\mnchanged{
A Seiberg-like dual pair of solitons was previously found for domain wall solutions \cite{Isozumi:2004jc}
and was then nicely understood in a D-brane configurations by
exchanging of positions of two NS5-branes along one direction
\cite{Eto:2004vy}.
}
}

In the end we expect the two dual vortex moduli spaces to be deformed and/or modified in the (respective) infinite gauge coupling
limits in such a way that they reduce to the same moduli space of Grassmannian lumps. Indeed, in the lump limit, the master equation~(\ref{mastereq}) can be solved algebraically by
\beq \label{lumpsol}
\Omega(z,\bar{z}) = \xi^{-1} \, H_0(z) H_0^\dagger(\bar{z}).
\eeq
$\det H_0 H_0^\dagger$ must be non-vanishing in order to have non-singular configurations. A set of
parameters for which $\det H_0 H_0^\dagger=0$ at some point on the $z$ plane corresponds to a small lump singularity which must
be discarded. In terms of ${\bf R}(z)$, such unphysical singularities can be avoided by means of the constraint (see Eq.~(\ref{h0condition}))
\begin{eqnarray}
\quad {}^\forall z:\quad |P(z)|^2\det\left({\bf 1}_{N_{\rm C}}+{\bf R}(z){\bf
R}^{\dagger}(z)\right)\neq0,\label{genresultant}
\end{eqnarray}
that is nothing but the generalization of Eq.~(\ref{resultant}), to which it correctly reduces for $\NC=1$ and $\NF=2$. This nicely completes the extension of the holomorphic rational map approach to Grassmannian lumps.

It is possible to show that the ``lump'' condition $\det H_0 H_0^\dagger \neq 0$ is equivalent to statement that
$({\bf Z,\widetilde \Psi})$ is $GL(k,{\bf C})$ free
(see Appendix \ref{Ohashi}), so that the moduli space of $k$-lumps is given by:
\begin{eqnarray}
\nonumber {\cal M}^{{\rm lump}}_{\NC,\NF;k} & = & \left\{ ({\bf Z,\Psi,\widetilde \Psi}): GL(k,{\bf C})\; {\rm free\; on}\; ({\bf
Z,\Psi})\; {\rm
and}\; ({\bf Z,\widetilde \Psi}) \right\}/GL(k,{\bf C}) \\
& = & {\cal M}_{\NC,\NF;k} \cap {\cal M}_{\tNC,\NF;k}.\label{lumpspace}
\end{eqnarray}
Namely, the moduli space of $k$-lumps is the intersection of the moduli space of $k$-vortices in one theory with
that of the Seiberg dual.
The physical interpretation is the same as for singularities in the ${\bf C}P^1$ lump moduli space. Increasing $g^2$ the semilocal
vortex moduli space is deformed and approaches that of Grassmannian lumps and, in the infinite coupling limit, it only develops
small lump singularities, as expected. These singularities correspond exactly to the presence of local
vortices, whose sizes, \mnchanged{$1/g\sqrt \xi$}, shrink to zero in the infinite gauge coupling limit. The same occurs in the Seiberg dual theory. What is
left after the removal of the singular points is nothing but the intersection of the two vortex moduli spaces. In other words, the moduli space of semilocal vortices in each dual theory is given by the {same} moduli space of lumps
 in which we ``blow-up'' the small lump singularities with the
insertion of the local vortex moduli subspace of the respective theory. From these considerations it is easy to convince ourselves that
(\ref{lumpspace}) is correct: taking the intersection in
(\ref{lumpspace}) eliminates the local vortices of both dual theories, leaving us with the moduli space of lumps.

The moduli space of semilocal strings has also been constructed, in terms of the symplectic quotient (\ref{theorem2}),
using a D-brane setup \cite{Hanany:2003hp}. In this approach one must identify the parameter $r$ with the gauge coupling
of the four dimensional gauge theory:
\beq
r={2 \pi } / g^2.
\eeq
As one may expect, the lump limit is seen to be formally equivalent to taking the limit $r \rightarrow 0$. This limit is
singular, in fact it develops singularities that correspond to the already mentioned small lump singularities. We can write for
the moduli space of lumps:
\beq {\cal M}^{{\rm lump}}_{\NC,\NF;k}=\{({\bf Z}, {\bf \Psi}, \tilde{\bf \Psi}) :D=[{\bf Z}^\dagger,{\bf Z}]+ {\bf \Psi}^\dagger {\bf \Psi} -{\bf \tilde{\Psi}}{\bf \tilde{\Psi}}^{\dagger}=0, \ U(k) \hbox{ free }\}
/U(k), \label{symplecticlumps}
\eeq
were we have excluded ``by hand" the small lump singularities \mnchanged{by} considering only points for which the $U(k)$ action is free\footnote{A free quotient of a compact group is always smooth.}.

 Coming back to our example $W{\bf C}P^1_{(1,-1)}$, from Section~\ref{msq}, we see that we must take away both of the
pathological points that spoil the Hausdorff property, instead of only one. The net result is the intersection of the two
regularized spaces $W{\bf C}P^1_{(\underline{1},-1)}$ and $W{\bf C}P^1_{(1,\underline{-1})}$, e.g. ${\bf C}^*$. This is also the
space that we obtain if we eliminate the singularity of the conifold, in agreement with the general statement Eq.~(\ref{symplecticlumps}).

The moduli space duality and the lump limit are then summarized by the following ``diamond'' diagram:
$$
\xymatrix{ & \hat{\cal M}_{\NC,\NF;k} \ar[dr] \ar[dl] & \\
{\cal M}_{\NC,\NF;k} \ar[dr]_{g^2\to\infty} & \ar@{.>}[r] \hbox{\small Seiberg duality} \ar@{.>}[l]& {\cal M}_{\tNC,\NF;k} \ar[dl]^{\tilde{g}^2 \to \infty} \\
& {\cal M}^{{\rm lump}}_{\NC,\NF;k} & }
$$

\section{Some examples} \label{example}

\subsection{Fundamental semilocal vortices and lumps} \label{single}

In this section we consider the topological sector $k=1$, which consists of fundamental (single) semilocal vortices and lumps. The
basic mathematical objects in this case are the weighted projective spaces with both positive and negative weights (see
Appendix~\ref{wpsnw}). For these we adopt the notation $W{\bf C}P^{n-1}[Q_1^{w_1},\ldots,Q_{l}^{w_{l}}]$, where the $Q_i$,
$i=1,\ldots,l(\le n)$, represents the weight and $w_i$ the number of homogeneous coordinates carrying that weight; clearly
$\sum_{i=1}^l w_i=n$.

This particular kind of toric variety plays a fundamental role in gauged sigma models in two dimensions \cite{Witten:1993yc} and
their solitons \cite{Schroers:1996zy}. As was noted in \cite{Witten:1993yc}, when the set of weights includes both positive and
negative integers (recall that multiplying all of the weights by a common integer number has no effect), the space is non-Hausdorff.
There are two possible regularizations (in the sense of Section~\ref{msq}), which correspond to eliminating
\nychanged{the subspace} where either all
positively charged or all negatively charged coordinates vanish.

Looking at Eq.~(\ref{GLaction}), it is easy to see that ${\bf Z}$ ``decouples'', in the sense that the $GL(1,{\bf C})=U(1)^{\bf
C}={\bf C}^*$ acts trivially on it; indeed
\beq \left({\bf Z}, {\bf \Psi}, \tilde{\bf \Psi}\right) \sim \left({\bf Z}, \lambda^{-1} {\bf \Psi}, \lambda\tilde{\bf \Psi}\right), \qquad \lambda \in {\bf C}^*.
\label{bee}\eeq
Although we shall keep the same notation for the moduli spaces, they will be intended as the internal moduli spaces from now on, as the
position moduli ${\bf Z}\in {\bf C}$ always factorize. Given this, we identify $\hat{{\cal M}}_{\NC,\NF;1}$ with $W{\bf
C}P^{\NF-1}[1^{\NC},-1^{\tNC}]$. We can regularize (in the sense of Section~\ref{msq}) this space by insisting that ${\bf \Psi} \neq 0$. We
indicate this space with the following notation:
\beq{\cal M}_{\NC,\NF;1} = {\cal O}(-1)^{\oplus\tNC} \rightarrow {\bf C}P^{\NC-1} \equiv W{\bf
C}P^{\NF-1}[\underline{1^{\NC}},-1^{\tNC}]
\eeq
where ${\cal O}(-1) $ stands for the universal line bundle\footnote{The fiber of the universal line bundle at each
point in ${\bf C}P^{n-1}$ is the line that it represents in ${\bf C}^{n}$.}.
Analogously, the dual regularization is obtained by imposing $\tilde{\bf \Psi} \neq 0$:
\beq{\cal M}_{\tNC,\NF;1} = {\cal O}(-1)^{\oplus\NC} \rightarrow {\bf C}P^{\tNC-1} \equiv W{\bf C}P^{\NF-1}[\underline{1^{\tNC}},
{-1^{\NC}}].
\eeq
Note that when $\NC=\tNC$ these spaces become
\mnchanged{non-compact (local)} Calabi-Yau manifolds, which corresponds to the fact that just in this case the conformal bound for a four-dimensional $U(\NC)$ with ${\cal N}=2$ supersymmetry is saturated.

In the lump limit one must take ${\bf \Psi},\tilde{\bf \Psi} \neq 0$:
\beq {\cal M}^{{\rm lump}}_{\NC,\NF;1}=({\bf C}^{\tNC})^* \ltimes {\bf C}P^{\NC-1} \simeq ({\bf C}^{\NC})^* \ltimes {\bf C}P^{\tNC-1} \simeq \{({\bf C}^{\NC})^* \oplus ({\bf C}^{\tNC})^*\}/ {\bf C}^*,
\label{lumpsym} \eeq
\mnchanged{ with $F \ltimes B$ denoting a fiber bundle with $F$ a fiber and $B$ a base. } The ${\bf C}^*$ acts with
charges $+1$ and $-1$ on $({\bf C}^{\NC})^*$ and $({\bf C}^{\tNC})^*$ respectively. If we define
\beq
{\cal M}^{{\rm lump}}_{\NC,\NF;1} \equiv W{\bf C}P^{\NF-1}[\underline{1^{\NC}},\underline{-1^{\tNC}}],
\eeq
we can summarize the situation with the following diamond diagram:
$$
\xymatrix{ & W{\bf C}P^{\NF-1}[1^{\NC},-1^{\tNC}] \ar[dr] \ar[dl] & \\
W{\bf C}P^{\NF-1}[\underline{1^{\NC}},{-1^{\tNC}}] \ar[dr]_{g^2\to\infty} & \ar@{.>}[r] \hbox{\small Seiberg duality} \ar@{.>}[l]& W{\bf C}P^{\NF-1}[\underline{1^{\tNC}},-1^{\NC}] \ar[dl]^{\tilde{g}^2 \to \infty} \\
& W{\bf C}P^{\NF-1}[\underline{1^{\NC}},\underline{-1^{\tNC}}] & }
$$

Let us consider some concrete examples.
\begin{itemize}
\item \underline{$\NC=1, \NF=2$}

This is a self-dual system. From $\hat{{\cal M}}_{1,2;1}=W{\bf C}P^1[1,-1]$
one finds ${\cal M}_{1,2;1}={\bf C}$ for both the original
and the dual theory. The moduli space of lumps is obtained by removing the small lump singularity and it is ${\cal
M}^{{\rm lump}}_{1,2;1}={\bf C}^*$. Explicitly, from the moduli matrix
\beq H_0=(z-z_0,b)
\quad \Leftrightarrow \quad \left\{ {\bf Z},{\bf \Psi},\tilde{\bf \Psi} \right\} = \left\{ z_0,1,b \right\},
\eeq
one finds the solution~(\ref{lumpsol})
\beq \Omega=|z-z_0|^2+|b|^2
\eeq
and the non-vanishing condition is $b\neq 0$ (consider the point $z=z_0$). Removing the point $b=0$ from the vortex
moduli space ${\bf C}$ one obtains the lump moduli space ${\bf C}^*$. In summary:
$$
\xymatrix{ & W{\bf C}P^1[1,-1] \ar[dr] \ar[dl] & \\
{\bf C} \ar[dr]_{g^2\to\infty} & \ar@{.>}[r] \hbox{\small Seiberg duality} \ar@{.>}[l]& {\bf C} \ar[dl]^{\tilde{g}^2 \to \infty} \\
& {\bf C}^* & }
$$

\item \underline{$\NC=2, \NF=3$ dual to $\NC=1, \NF=3$}

We have now the ``parent'' moduli space, $\hat{{\cal M}}_{2,3;1}=W{\bf C}P^2[1,1,-1]$.
The two dual regularizations are ${\cal M}_{2,3;1}=\tilde{{\bf C}}^2$,
namely the blow-up of the origin of ${\bf C}^2$ by inserting $S^2 \simeq {\bf C}P^1$
(see Appendix~\ref{blow up}), and ${\cal M}_{1,3;1}={\bf C}^2$. The lump limit is
${\cal M}_{2,3;1}^{{\rm lump}}=({\bf C}^2)^*$, the two-dimensional complex vector space minus the origin.

All of the moduli spaces \nychanged{can} be found using the moduli matrix.
\nychanged{In the lump limit, the general solution for the original theory leads}
\beq H_0=\begin{pmatrix}
1 & b & 0 \cr 0 & z-z_0 & c
\end{pmatrix}
\quad \Leftrightarrow \quad \left\{ {\bf Z},{\bf \Psi},\tilde{\bf \Psi} \right\} = \left\{
z_0,\left(\begin{array}{c}-b\\1\end{array}\right),c \right\}, \label{moduli23} \eeq
\beq \det \Omega \rvert_{z=z_0} = |c|^2(1+ |b|^2)
\eeq
and so the determinant vanishes, indicating a small lump singularity, on the blown-up 2-sphere $c=0$. Removing this 2-sphere
from the vortex moduli space $\tilde{{\bf C}}^2$ one is left with the lump moduli space $({\bf C}^2)^*$.
In order to cover the whole moduli space $\tilde {\bf C}^2$, together with
that in Eq.~(\ref{moduli23}), one needs another patch for the moduli matrix. The transition functions between these two patches are given in Appendix~\ref{blow up}.
In the case of the dual theory
\beq H_0=(z-\tilde{z}_0,\tilde{b},\tilde{c})
\quad \Leftrightarrow \quad \left\{ {\bf Z},{\bf \Psi},\tilde{\bf \Psi} \right\} = \left\{ \tilde z_0,1,\left(\tilde
b,\tilde c\right) \right\}, \label{matbelian}
\eeq
\beq \Omega \rvert_{z=\tilde{z_0}}=|\tilde{b}|^2+|\tilde{c}|^2
\label{detabelian}
\eeq and so the determinant vanishes at the point $\{\tilde{b}=\tilde{c}=0\}\in{\bf C}^2$. The diamond diagram is
$$
\xymatrix{ & W{\bf C}P^2[1,1,-1] \ar[dr] \ar[dl] & \\
\tilde{{\bf C}}^2 \ar[dr]_{g^2\to\infty} & \ar@{.>}[r] \hbox{\small Seiberg duality} \ar@{.>}[l]& {\bf C}^2 \ar[dl]^{\tilde{g}^2 \to \infty} \\
& ({\bf C}^2)^* & }
$$

\item \underline{$\NC$, $\NF=\NC+1$ dual to $\NC=1,\NF$}

This is a generalization of the previous two examples. The parent space is $\hat{\cal M}_{\NC,\NC+1;1}=W{\bf
C}P^{\NC}[1^{\NC},-1]$. On one side we have ${\cal M}_{\NC,\NC+1;1}=W{\bf C}P^{\NC}[1^{\NC},\underline{-1}]=\tilde{\bf C}^{\NC}$,
which is ${\bf C}^{\NC}$ with the origin blown up into a ${\bf C}P^{\NC-1}$, while on the other side the dual moduli space is
simply ${\cal M}_{1,\NC+1;1}=W{\bf C}P^{\NC}[\underline{1^{\NC}},-1]={\bf C}^{\NC}$. In the lump limit we are left with ${\cal
M}_{\NC,\NC+1;1}^{{\rm lump}}=({\bf C}^{\NC})^*$ since in the original theory
\beq H_0=\begin{pmatrix}
{\bf 1}_{\NC-1} & {\bf b} & 0 \cr 0 & z-z_0 & c
\end{pmatrix}
\quad \Leftrightarrow \quad \left\{ {\bf Z},{\bf \Psi},\tilde{\bf \Psi} \right\} = \left\{
z_0,\left(\begin{array}{c}-{\bf b}\\1\end{array}\right),c \right\},\label{nonabmoduli}
\eeq
\beq \det \Omega \rvert_{z=z_0} = |c|^2(1+ |{\bf b}|^2)
\eeq
and so the small lump singularity is the blown-up ${\bf C}P^{\NC-1}$ at $c=0$ in the vortex moduli space.
Here ${\bf b}$ is a column ($\NC-1$)-vector.
For the dual theory
\beq H_0=(z-\tilde{z}_0,{\bf \tilde{b}})
\quad \Leftrightarrow \quad \left\{ {\bf Z},{\bf \Psi},\tilde{\bf \Psi} \right\} = \left\{ \tilde z_0, 1, \tilde{\bf
b} \right\},
\eeq
\beq
\Omega \rvert_{z=\tilde{z}_0}=|{\bf \tilde{b}}|^2
\eeq
which identifies the small lump singularity with the point $|{\bf \tilde{b}}|=0 \in {\bf C}^{\NC}$.
Here ${\bf \tilde b}$ is a row $\NC$-vector.
These moduli spaces are summarized by
the diamond diagram
$$
\xymatrix{ & W{\bf C}P^{\NC}[1^{\NC},-1] \ar[dr] \ar[dl] & \\
\tilde{{\bf C}}^{\NC} \ar[dr]_{g^2\to\infty} & \ar@{.>}[r] \hbox{\small Seiberg duality} \ar@{.>}[l]& {\bf C}^{\NC} \ar[dl]^{\tilde{g}^2 \to \infty} \\
& ({\bf C}^{\NC})^* & }
$$

\item \underline{$\NC=2, \NF=4$}

This theory is again self-dual. The parent space is $\hat{\cal M}_{2,4;1}=W{\bf C}P^3[1,1,-1,-1]$, which yields ${\cal
M}_{2,4;1}={\cal O}(-1) \oplus {\cal O}(-1)\rightarrow {\bf C}P^1$, namely the resolved conifold \mnchanged{\cite{Candelas:1989js}} (see Appendix~\ref{rescon}). The
moduli space of lumps is ${\cal M}_{2,4;1}^{{\rm lump}}=({\bf C}^2)^*\ltimes {\bf C}P^1$. Indeed
\beq H_0=\begin{pmatrix}
1 & b & 0 & 0 \cr 0 & z-z_0 & c & d
\end{pmatrix}
\quad \Leftrightarrow \quad \left\{ {\bf Z},{\bf \Psi},\tilde{\bf \Psi} \right\} = \left\{
z_0,\left(\begin{array}{c}-b\\1\end{array}\right),\left(c,d\right) \right\},
\eeq
and from the non-vanishing condition
\beq \det \Omega \rvert_{z=z_0} = (1+ |b|^2)(|c|^2+|d|^2) =|{\bf \Psi}|^2|{\bf \tilde{\Psi}}|^2\neq 0
\eeq
we recognize $(c,d)$ as coordinates of $({\bf C}^2)^*$ and $b$ as the inhomogeneous coordinate of the
base ${\bf C}P^1$. Therefore one removes the ${\bf C}P^1$ at $c=d=0$, that is ${\bf \tilde{\Psi}}=0$. In the dual theory, on the other
hand, the roles of ${\bf \Psi}$ and ${\bf \tilde{\Psi}}$ are interchanged and so one instead removes the ${\bf C}P^1$ at ${\bf
\Psi}=0$, which is related by a flop transition to the ${\bf C}P^1$ of the previous moduli space. In the end
$$
\xymatrix{ & W{\bf C}P^3[1,1,-1,-1] \ar[dr] \ar[dl] & \\
{\cal O}(-1) \oplus {\cal O}(-1)\rightarrow {\bf C}P^1
\ar[dr]_{g^2\to\infty} & \ar@{.>}[r] \hbox{\small Seiberg duality} \ar@{.>}[l]& {\cal O}(-1) \oplus {\cal O}(-1)\rightarrow {\bf C}P^1 \ar[dl]^{\tilde{g}^2 \to \infty} \\
& ({\bf C}^2)^*\ltimes {\bf C}P^1 & }
$$
\end{itemize}

It is suggestive to note that similar topological transitions of the type descried above occur within the non-commutative vortex moduli space as the non-commutativity parameter is varied \cite{Hanany:2003hp}.

\subsection{Multiple semilocal vortices and lumps} \label{double}

Let us now consider configurations with several vortices (or, equivalently, higher winding number vortices). Consider first the situation when all vortices are separated. In this case
the moduli space reduces to the symmetric product of single vortex moduli spaces \cite{Eto:2005yh}:
\beq
{\cal M}_{\NC,\NF;k }\big|_{sep} \simeq ({\cal M}_{\NC,\NF;1})^k / \mathfrak{S}_k, \label{sepvort}
\eeq where $\mathfrak{S}_k$ is the permutation group of $k$ objects. From this we can easily generalize our picture of the duality:

$$
\xymatrix{ & (\hat{\cal M}_{\NC,\NF;1})^k / \mathfrak{S}_k \ar[dr] \ar[dl] & \\
({\cal M}_{\NC,\NF;1})^k / \mathfrak{S}_k \ar[dr]_{g^2\to\infty} & \ar@{.>}[r] \hbox{\small Seiberg duality} \ar@{.>}[l]& ({\cal M}_{\tNC,\NF;1})^k / \mathfrak{S}_k \ar[dl]^{\tilde{g}^2 \to \infty} \\
& ({\cal M}^{{\rm lump}}_{\NC,\NF;1})^k / \mathfrak{S}_k & }
$$

It is well known that (\ref{sepvort}) contains orbifold singularities (which are resolved in the complete space)
that correspond to two or more coincident vortices. In order to see how duality works in this case, we can study in detail the moduli (sub)space of two coincident
vortices, generalizing the analysis of \cite{Eto:2006cx} to the semilocal case.

\mnchanged{
Let us restrict ourselves to double vortices ($k=2$).
}
The ``parent" space
\mnchanged{of coincident two vortices}
is found to be a kind of weighted Grassmannian manifold with negative weights (see Appendix~\ref{compsemvort}):
\beq
\hat {\cal M}_{\NC,\NF;2}\big|_{{\rm coinc}} = WGr_{\NC + \tNC + 1,2}^{(1^{\NC},0,-1^{\tNC})} = WGr_{\NF + 1,2}^{(1^{\NC},0,-1^{\tNC})}.
\label{coincsemiloc} \eeq
This space suffers from the same problems of regularization as weighted projective spaces with negative weights. We can regularize
it in two different (and dual) ways. The first is to choose only the points such that the first $2 \times (\NC +1)$ minor of the matrix defining the
$WGr_{\NC + \tNC + 1,2}^{(1^{\NC},0,-1^{\tNC})}$ is of rank 2 (see the definition of the matrix $M$ in Eq.~(\ref{semilocalset})). This gives us the moduli space for the theory
with $\NC$ colors, that we indicate with the following notation:
\beq
 {\cal M}_{\NC,\NF;2}\big|_{{\rm coinc}} \equiv WGr_{\NF + 1,2}^{(\underline{1^{\NC}},0,-1^{\tNC})}. \label{coincsemiloc1}
\eeq
The second possibility is to choose only the points such that the last $2 \times (\tNC +1)$ minor is of rank 2. This gives us the
moduli space for the dual theory with $\tNC$ colors:
\beq
 {\cal M}_{\tNC,\NF;2}\big|_{{\rm coinc}} \equiv WGr_{\NF + 1,2}^{(1^{\NC},0,\underline{-1^{\tNC}})}. \label{coincsemiloc2}
\eeq
These spaces are Calabi-Yau when $\NC^2=\tNC$ (see Appendix \ref{compsemvort}).

The moduli space of lumps is obtained by considering the intersection of the two dual spaces:
\beq
 {\cal M}^{{\rm lump}}_{\NC,\NF;2}\big|_{{\rm coinc}} = WGr_{\NF + 1,2}^{(\underline{1^{\NC}},0,\underline{-1^{\tNC}})}, \label{coincsemiloc3}
\eeq
where the two underlines mean that the first $2 \times (\NC+1)$ and the last $2 \times (\tNC+1)$ minors are both of rank 2.

We summarize this situation with the following diagram:
$$
\xymatrix{ & WGr_{\NF + 1,2}^{(1^{\NC},0,-1^{\tNC})} \ar[dr] \ar[dl] & \\
WGr_{\NF + 1,2}^{(\underline{1^{\NC}},0,-1^{\tNC})} \ar[dr]_{g^2\to\infty} & \ar@{.>}[r] \hbox{\small Seiberg duality} \ar@{.>}[l]& WGr_{\NF + 1,2}^{(1^{\NC},0,\underline{-1^{\tNC}})} \ar[dl]^{\tilde{g}^2 \to \infty} \\
& WGr_{\NF + 1,2}^{(\underline{1^{\NC}},0,\underline{-1^{\tNC}})} & }
$$
Let us consider an explicit example:

\begin{itemize}
  \item \underline{$\NC=2, \NF=3$ dual to $\NC=1, \NF=3$}
\end{itemize}

In this case we have $\hat {\cal M}_{2,3;2}\big|_{{\rm coinc}} \equiv WGr_{4,2}^{(1,1,0,-1)}$. This space, though the simplest example of non-abelian multiple semilocal vortex, already has a quite complicated structure (see Appendix~\ref{compsemvort}). It is not difficult to find the moduli space for the dual abelian theory (recovering the well-known result of \cite{Gibbons:1992gt}):
\beq {\cal M}_{1,3;2}\big\vert_{{\rm coinc}} \equiv WGr_{4,2}^{({1,1},0,\underline{-1})}={\bf C}^4, \label{abelian}\eeq
The non-abelian case turns out to be (Appendix~\ref{compsemvort}) the blow-up of a conifold embedded in ${\bf C}^5/\mathbb{Z}_2$. The action of $\mathbb{Z}_2$ on ${\bf C}^5(x_1,x_2,x_3,x_4,x_5)$ is $x_1=-x_1$, while the blow up of ${\bf
C}^5/\mathbb{Z}_2$ must be done along the subspace $x_1=x_2=x_3=0$. The conifold can be described by the algebraic equation
\beq
x^2_1-x_4 x_2+ x_5 x_3=0.
\eeq
Thus we can write
\beq
 {\cal M}_{2,3;2}\big\vert_{{\rm coinc}} = WGr_{4,2}^{(\underline{1,1},0,-1)}= \{ x^2_1-x_4 x_2+ x_5 x_3=0 \subset \tilde {\bf C}^5 / \mathbb{Z}_2 \}.
\label{cone1}
\eeq
It is interesting to see how the space (\ref{cone1}) reduces, in the lump limit, to the same space that follows from (\ref{abelian}):
\beq
{\cal M}^{{\rm lump}}_{2,3;2}\big\vert_{{\rm coinc}} \equiv WGr_{4,2}^{(\underline{1,1},0,\underline{-1})}=({\bf C}^2)^* \times {\bf C}^2.
\eeq
We summarize the duality relations for this example:
$$
\xymatrix{ & WGr_{\NF + 1,2}^{(1^{\NC},0,-1^{\tNC})} \ar[dr] \ar[dl] & \\
 x^2_1-x_4 x_2+ x_5 x_3=0 \subset \tilde {\bf C}^5 / \mathbb{Z}_2 \ar[dr]_{g^2\to\infty} & \ar@{.>}[r] \hbox{\small Seiberg duality} \ar@{.>}[l]& {\bf C}^4 \ar[dl]^{\tilde{g}^2 \to \infty} \\
& ({\bf C}^2)^* \times {\bf C}^2. & }
$$
\section{Normalizability of zero-modes and the effective action} \label{effact}
The effective theory on the vortex worldsheet is obtained via the usual procedure \mnchanged{\cite{Manton:1981mp}} of promoting the moduli to slowly varying fields
on the worldsheet \cite{Auzzi:2003fs,Eto:2006pg,Shifman:2006kd}. It turns out to be a two dimensional sigma model whose K\"ahler
potential can be calculated from the moduli matrix \cite{Eto:2006pg,Eto:2006uw}
:
\begin{eqnarray}
K = {\rm Tr}\int\!\! d^2z \left( \xi \log\Omega+\Omega^{-1} H_0 H_0^\dagger + {\cal O}(1/g^2) \right). \label{kahler}
\end{eqnarray}
\mnchanged{
This formula with explicit expression of the third term was first obtained after tedious calculation in terms of component fields \cite{Eto:2006pg},
but the derivation has been drastically simplified by using superfields \cite{Eto:2006uw}.
}
By virtue of translational symmetry, it is possible to show that the center-of mass parameter is always decoupled
\cite{Eto:2006db} and, specifically, it appears with an ordinary kinetic term whose coefficient is proportional to the total
tension. The center-of-mass is a free field.

Let us concentrate on the other moduli. Analyzing the divergences of the K\"ahler potential, one can establish which
moduli \mnchanged{among $2k\NF$} have an infinite kinetic term in the Lagrangian and are non-normalizable. Fluctuations
of those moduli are frozen, as well as their motion in the geodesic approximation, while the evolution of the rest
of the moduli will be allowed. Very recently some evidence has been found that all modes become
normalizable when  semilocal vortices are coupled to gravity \cite{Aldrovandi:2007bn}.

\subsection{Non-normalizable modes}
The divergent terms of the K\"ahler potential can
\kochanged{come only from integrations
around the boundary $|z|=L$ ($L$ is a suitable infra-red cut-off),
since $\Omega$ is assumed to be invertible and
smooth.
Remembering that \nychanged{$\Omega \to \xi^{-1}H_0H_0^\dagger$} for large $z$, the
divergent terms can
be calculated keeping only the first term in Eq.~(\ref{kahler})}:
\begin{eqnarray}
\nychanged{\xi} \int^{|z|=L} d^2z \log\det (H_0H_0^\dagger)&\sim& \nychanged{\xi} \int^{|z|=L} d^2z \log\det
\left(\mathbf{D}^{-1}H_0(\mathbf{D}^{-1}H_0)^\dagger\right)\nn &=& \nychanged{\xi} \int^{|z|=L} d^2 z
\log\det \left({\bf 1}_{N_{\rm C}}
+\left|{\bf \Psi}\frac{{ 1}}{z-{\bf Z}}\widetilde {\bf \Psi}\right|^2\right)\nn &=&
\nychanged{\xi} \int^{|z|=L} d^2z\ \left[ \frac1{|z|^2}{\rm
Tr}\left|{\bf \Psi} \widetilde {\bf \Psi}\right|^2+{\cal O}(|z|^{-3}) \right] \nn
&=&\nychanged{2\pi\xi}\, \log L \, {\rm Tr}\left|{\bf \Psi} \widetilde {\bf
\Psi}\right|^2 +{\rm const.}+{\cal O}(L^{-1})\label{divergent}
\end{eqnarray}
where we used Eq.~(\ref{eq:DH}) and a K\"ahler transformation $K\to K+f+f^*$, with $f=\nychanged{\xi} \int d^2z \log\det {\bf D}^{-1}$. Equation
(\ref{divergent}) means that the elements of ${\bf \Psi} \widetilde {\bf \Psi}$ are non-normalizable and should be fixed. The
number of non-normalizable parameters crucially depends on the rank of ${\bf \Psi} \widetilde {\bf \Psi}$:
\begin{eqnarray}
r\equiv {\rm rank}\left({\bf \Psi} \widetilde {\bf \Psi}\right)\le{\rm min}\left(k,N_{\rm C},\tilde N_{\rm C}\right) \equiv j.
\label{inequality}
\end{eqnarray}
In the following we calculate the number of non-normalizable moduli when the above inequality is saturated, $r=j$. This happens for generic points of the moduli space. It follows
that for particular submanifolds of the moduli space when $r<j$ {\it the number of normalizable parameters is enhanced.} We will give a
simple example in Section~\ref{modulienhan}.

Using the global symmetry $SU(N_{\rm C})_{\rm C+F}\times SU(\tilde N_{\rm C})_{\rm F}$, we can always fix ${\bf \Psi}
\widetilde {\bf \Psi}$ to have the following form:
\begin{eqnarray}
{\bf \Psi}
\widetilde {\bf \Psi}=\left(
\begin{array}{cc}
\Lambda_r & {\bf 0}\\ {\bf 0}& {\bf 0}
\end{array}
\right)\label{standardpsitildepsi}
\end{eqnarray}
where $\Lambda_r={\rm diag}(\lambda_1,\cdots,\lambda_r)$ with positive real parameters $\lambda_i>0$. Note that this
symmetry of the vacuum is generally broken by the vortex and so it generates moduli for our solution. But the corresponding
moduli are non-normalizable, so that we will not count them in the following.

 To proceed further we have to distinguish two
cases:
\begin{itemize}
  \item $k\le {\rm min}(N_{\rm C},\tilde N_{\rm C})$

In this case the saturation of the inequality (\ref{inequality}) means $r=k$, and the matrices ${\bf \Psi}$ and $\widetilde {\bf
\Psi}$ have the following block-wise form (suffixes indicate dimensions of blocks):
\begin{eqnarray}
{\bf \Psi}=\left(
\begin{array}{c}
A_{[k \times k]} \\ B_{[(\NC-k) \times k]}
\end{array}
\right),\ \ \widetilde {\bf \Psi}=(C_{[k\times k]},D_{[k \times (\tilde \NC -k)]}),
\end{eqnarray}
from which we find $AC=\Lambda_k$. Because ${\rm det }\Lambda_k\not=0 \Rightarrow {\rm det}A\not=0$, we can completely fix
$GL(k,{\bf C})$ by taking $A={\bf 1}_k$. Thus, we obtain:
\begin{eqnarray}
{\bf \Psi}=\left(
\begin{array}{c}
{\bf 1}_k \\ {\bf 0}
\end{array}
\right), \quad \widetilde {\bf \Psi}=(\Lambda_k,{\bf 0}).
\end{eqnarray}
The corresponding moduli matrix is:
\begin{eqnarray}
H_0=\left(
\begin{array}{cccc}
z{\bf 1}_k-{\bf Z}&{\bf 0}&\Lambda_k&{\bf 0}\\
{\bf 0}&{\bf 1}_{N_{\rm C}-k}&{\bf 0}&{\bf 0}
\end{array}
\right).
\end{eqnarray}
From the above we find that the normalizable moduli are all contained in the $k \times k$ matrix ${\bf Z}$, so that:
\begin{eqnarray}
{\rm dim}{\cal M}_{N_{\rm C}, N_{\rm F};k}^{\rm norm}= 2 k^2.\label{normkleqnc}
\end{eqnarray}

From here it is easy to see that fundamental semilocal vortices, $k=1$, always have only 2 real moduli, corresponding to the
position on the plane. Orientation moduli are instead non-normalizable, independently of $\NC$ and $\NF$. This behavior is very
different from the local case, $\NC=\NF$.

\item $k\ge {\rm min}(N_{\rm C},\tilde N_{\rm C})$

  We assume $N_{\rm C}\le \tilde N_{\rm C}$ \mnchanged{without loss of generality} (The results for
$N_{\rm C}\ge \tilde N_{\rm C}$ are obtained using Seiberg duality $N_{\rm C}\leftrightarrow \tilde N_{\rm C}$).
Thus the saturation of Eq.~(\ref{inequality}) leads $k=N_{\rm C}$.
For ${\bf \Psi}$ and $\widetilde {\bf \Psi}$ we have the following block form:
\begin{eqnarray}
{\bf \Psi}=\left(A_{[\NC \times \NC]},B_{[\NC \times (k - \NC)]}\right), \widetilde {\bf \Psi}=\left(
\begin{array}{cc}
C_{[\NC \times \NC]}&D_{[\NC \times (\tilde \NC - \NC)]} \\ E_{[(k-\NC) \times \NC]} & F_{[(k-\NC) \times (\tilde\NC-\NC)]}
\end{array}\right).
\end{eqnarray}
We see that ${\bf \Psi}$ must have rank equal to $\NC$ so that can be always put in the following form:
\begin{eqnarray}
{\bf \Psi}=\left({\bf 1}_{N_{\bf C}},{\bf 0}\right)
\end{eqnarray}
via a $GL(k,{\rm C})$ transformation. Thus $\widetilde {\bf \Psi}$ and the remaining $GL(k,{\bf C})$ symmetry are
\begin{eqnarray}
\widetilde {\bf \Psi}=\left(
\begin{array}{cc}
\Lambda_{N_{\rm C}}&{\bf 0} \\ E&F
\end{array}\right),\quad
\left(
\begin{array}{cc}
{\bf 1}_{N_{\rm C}}&{\bf 0} \\ G&H
\end{array}\right)\in GL(k,{\bf C}).
\end{eqnarray}
Here $E$ can be fixed to be zero using $G$. We obtain:
\begin{eqnarray}
{\bf \Psi}=\left({\bf 1}_{N_{\rm C}},{\bf 0}\right), \quad \widetilde {\bf \Psi}=\left(
\begin{array}{cc}
\Lambda_{N_{\rm C}}&{\bf 0} \\ {\bf 0}&T
\end{array}\right), \quad
{\bf Z}=\left(
\begin{array}{cc}
X& Y \\ \tilde Y& W
\end{array}\right),
\end{eqnarray}
with the remaining $u\in GL(k-N_{\rm C},{\bf C})$ action:
\begin{eqnarray}
X\rightarrow X,\quad Y\rightarrow Yu, \quad \tilde Y\rightarrow u^{-1} \tilde Y,\quad W\rightarrow u^{-1}W u,\quad T\rightarrow
u^{-1} T.
\end{eqnarray}
The normalizable parameters are contained in the ${\bf Z}$ and $T$ matrices, from which we subtract the $(k-N_{\rm C})^2$
parameters of the remaining $GL$ action:
\begin{eqnarray}
{\rm dim}{\cal M}_{N_{\rm C}, N_{\rm F};k}^{\rm norm}&=& 2 \left(k^2+(k-N_{\rm C})(\tilde N_{\rm C}-N_{\rm C})-(k-N_{\rm
C})^2\right)\nn &=&2\left((N_{\rm C}+\tilde N_{\rm C})k-N_{\rm C}\tilde N_{\rm C}\right).\label{normkgeqnc}
\end{eqnarray}

The formula obtained is clearly symmetric in $\NC$ and $\tNC$. It is interesting to note, for instance, that \nychanged{$k=\NC$ vortices in
the $U(\NC)$ theory} always have $2\NC^2$ real, normalizable moduli for any $\NF$; they roughly correspond to the $2\NC$ positions in the
plane and the $2 \NC(\NC-1)$ relative sizes and orientations in the color-flavor space.

\end{itemize}

\subsection{Examples of enhancement of normalizable modes} \label{modulienhan}
\subsubsection{Abelian case}

In the abelian case we always have $k \ge \NC=1$. No submanifolds with an enhanced number of normalizable
moduli can be found. This is easy to understand directly from the moduli matrix
\beq H_0=\left(P(z), R_1(z),\ldots,R_{\tNC}(z)\right).
\eeq
Substituting this into the first line of Eq.~(\ref{divergent}) we can see that the only non-normalizable modes are the $\tNC$ coefficients of the leading power of the polynomials $R_i$. The ``local'' moduli in $P(z)$, associated to vortex positions, as well as the rest of the moduli in the semilocal part are instead normalizable:
\begin{eqnarray}
 {\cal M}_{1, N_{\rm F};k}^{\rm norm} ={\cal M}_{1,k}^{\rm local}\times {\bf C}^{(k-1)\tilde N_{\rm
 C}} ={\bf C}^{k+(k-1)(N_{\rm F}-1)}.
\end{eqnarray}

\subsubsection{Non-abelian cases}

\begin{itemize}
  \item $k=1$

A single vortex is characterized by its size $\Lambda_1=\lambda$. When $\lambda \neq 0$ we use the result (\ref{normkleqnc}) of
the previous section and conclude that the space of normalizable moduli is just given by the center of mass coordinates:
\beq
 {\cal M}_{N_{\rm C}, N_{\rm F};1}^{\rm norm}(\lambda\not=0)={\bf C}.
\eeq
The case $\lambda=0$ corresponds to a local vortex, so that:
\beq
{\cal M}_{N_{\rm C}, N_{\rm F};1}^{\rm norm}(\lambda=0)={\cal M}_{N_{\rm C};1}^{\rm local}={\bf C}\times {\bf C}P^{N_{\rm C}-1.}
\eeq
This is the simplest example of enhancement of normalizable moduli.
\mnchanged{The former corresponds to the situation discussed
by Shifman and Yung \cite{Shifman:2006kd}.}
Note that $\lambda$ is fixed in the dynamics of the vortex,
so it makes sense to consider different regimes at different values of this parameter. In the two cases above the effective low energy
theory is completely different.
In fact $\lambda\sim 1/g\sqrt \xi$ represents a transition region, in which the effective theory description must be appropriately changed due to an increased number of
massless degrees of freedom that develop for $\lambda\to 0$ (see also the discussion at the beginning of Section~\ref{dyna}).


 \item $k=2$

Next we consider configurations with two vortices. Now we have two size parameters $\Lambda_2={\rm diag }(\lambda_1,\lambda_2)$. When both
sizes do not vanish, $\lambda_1,\lambda_2\not =0$, which is the generic case, we use again (\ref{normkleqnc}) to obtain
\begin{eqnarray}
{\cal M}_{N_{\rm C},N_{\rm F};k=2}^{\rm norm} (\lambda_1,\lambda_2\not=0) ={\bf C}^{k^2}\big|_{k=2}={\bf C}^{4}.
\end{eqnarray}


In this case, the (normalizable) moduli space for semilocal vortices and for lumps are the same. Two out of four are moduli for positions and their fluctuations are localized around the corresponding vortex. The other two are for a relative
size and a relative orientation. Using the results for a single vortex, one sees that fluctuations of the latter two cannot be localized
around a vortex only, but should be localized around (between) the two vortices because of their normalizability (concrete examples are shown in Fig.~\ref{lumpwf}).
\begin{figure}[ht]
\begin{tabular}{ccc}
\includegraphics[width=5.3cm]{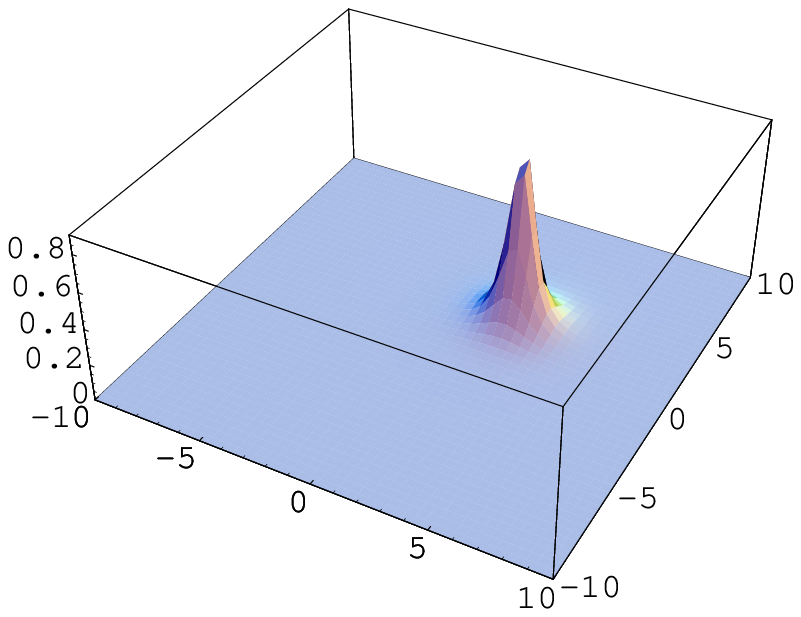} &
\includegraphics[width=5.3cm]{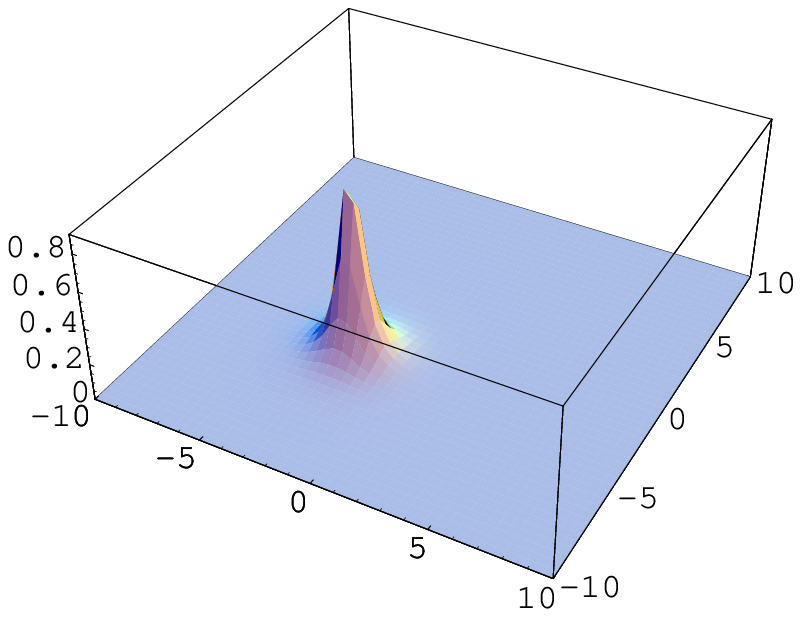} &
\includegraphics[width=5.3cm]{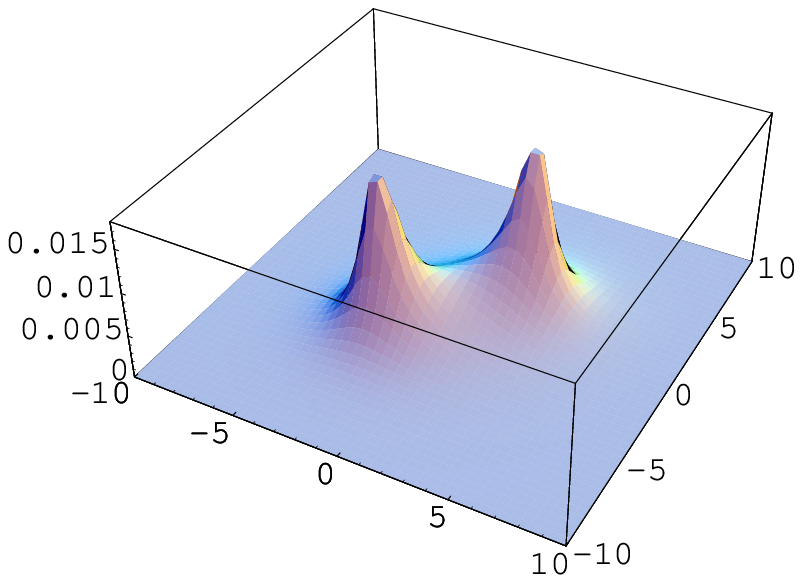} \\
{\small (a) Lump 1 position modulus}& {\small (b) Lump 2 position modulus} & {\small (c) Relative orientation}
\end{tabular}
\label{}
\caption{\small Wave functions of some normalizable moduli fields for $k=2$ lumps in $\NC=2$, $\NF=4$ model. Position moduli look localized around the corresponding vortex, while relative orientation modulus lies in between.} \label{lumpwf}
\end{figure}

In the opposite case, when both sizes vanish, $\lambda_1=\lambda_2=0$, we deal with a very fine-tuned point in the moduli space,
${\rm rank} \ \Lambda=0$, and we expect an enhancement of massless moduli.

To study this situation let us start for simplicity with $N_{\rm C}=2$. In the $(1,1)$ patch (of the type (\ref{11patch})), ${\bf
\Psi}^{(1,1)}={\bf 1}_2$, so that the vanishing size means ${\bf \tilde \Psi}^{(1,1)}=0$. In the $(0,2)$ patch (resp. the $(2,0)$ patch), of the type
(\ref{20patch}) (resp. (\ref{02patch})), this also imposes ${\bf \tilde \Psi}^{(2,0)}=0 \ ({\bf \tilde \Psi}^{(0,2)}=0$) except for the points on the subspace with $a=0
 \ (a'=0)$. Thus we obtain:
 \beq
 {\cal M}_{2, N_{\rm F};2}^{\rm norm}\supset {\cal M}_{2;2}^{\rm local},\quad {\cal M}_{2, N_{\rm F};2}^{\rm norm}\big|_{a \neq 0} = {\cal M}_{2;2}^{\rm local}\big|_{a \neq 0}
 \eeq

In the subspace with $a=0 $ and $a'=0$, $\bf \widetilde \Psi$ can take non-zero values:
\begin{eqnarray}
{\bf Z}=\left(
 \begin{array}{cc}
  0&1 \\ \beta&\alpha
 \end{array}\right),\quad
 {\bf \Psi}=\left(
 \begin{array}{cc}
  b_1&0 \\ b_2&0
 \end{array}\right),\quad
\widetilde{\bf \Psi}=\left(
 \begin{array}{ccc}
  0&\cdots &0\\ c_1 & \cdots& c_{\tNC}
 \end{array}\right)
\end{eqnarray}
where $(b_1,b_2)=(1,b)$ for the $(0,2)$ patch and $(b_1,b_2)=(b',1)$ for the $(2,0)$ patch. Under the remaining $GL(2,{\bf C})$
transformation $U=u {\bf 1}_2 \ (u \in {\bf C}^*)$, we find the following equivalence:
\begin{eqnarray}
(b_1,b_2,c_1,\cdots,c_{\tNC}) \sim (u\, b_1,u\,b_2,u^{-1}c_1,\cdots,u^{-1}c_{\tNC}),
\end{eqnarray}
so that these parameters define the space: \kochanged{$W{\bf C}P^{\tilde N_{\rm C}+1}[\underline{1^2}, -1^{\tilde N_{\rm C}}]$.}
We can easily generalize this result when the number of color is arbitrary: $W{\bf C}P^{N_{\rm F}-1}[ \underline{1^{N_{\rm
C}}},-1^{\tilde N_{\rm C}}]$. We have:
\beq
{\cal M}_{\NC,\tNC;2}^{\rm norm}\big|_{a = 0}=
{\bf C}^2 \times W{\bf C}P^{N_{\rm F}-1}[\underline{1^{N_{\rm C}}},
-1^{\tilde N_{\rm C}}].
\eeq
This result can be easily understood by noting that the $a=0$ case describes two parallel (in the color space) vortices, thus we reduce to the
case of a $k=1$ vortex: ${\cal M}_{\NC,\tNC;2}^{\rm norm}\big|_{a = 0}={\cal M}_{\NC,\tNC;1}$. A configuration of
this type can also be considered as \nychanged{an embedding of (a nontrivial bundle of)}
the abelian semilocal case $k=2$:
\begin{eqnarray}
 {\bf C}^2\times W{\bf C}P^{N_{\rm F}-1}[
\underline{1^{N_{\rm C}}},-1^{\tilde N_{\rm C}}] ={\cal M}_{1, N_{\rm
 F};2}^{\rm norm}\ltimes {\bf C}P^{N_{\rm C}-1}.
\end{eqnarray}
Finally we find:
\begin{eqnarray}
 {\cal M}_{N_{\rm C}, N_{\rm F};2}^{\rm norm}(\lambda_1=\lambda_2=0) = {\cal M}_{N_{\rm C};2}^{\rm local}\big|_{a \neq 0} \cup \left(
 {\cal M}_{1, N_{\rm F};2}^{\rm norm} \ltimes {\bf C}P^{N_{\rm C}-1} \right).
\end{eqnarray}

Moduli space of normalizable modes for lumps is obtained by removing ${\cal M}_{N_{\rm C};2}^{\rm local}$:
\begin{eqnarray}
 {\cal M}_{N_{\rm C}, N_{\rm F};2}^{{\rm lump}-{\rm norm}} (\lambda_1=\lambda_2=0) ={\bf C}^2\times W{\bf C}P^{N_{\rm
 F}-1}[\underline{1^{N_{\rm C}}},\underline{-1^{\tilde N_{\rm C}}}].
\end{eqnarray}

The mixed case with $\lambda_1\not =0, \lambda_2=0$ is more complex, and we will not treat it here.
\end{itemize}

\subsection{Dynamics of the effective  $k=1$ vortex  theory}
\label{dyna}

The presence of non-normalizable modes has a remarkable consequence in the low-energy effective description
of the vortex. As we have seen, these modes must be fixed, they are not dynamical. Even more remarkable are the
consequences of the presence of non-normalizable modes with the physical dimension of a length, such as the size
moduli. In this case the derivative expansion in the effective action will contain, generically, powers of $\lambda \,
\partial$, where $\lambda$ is the size moduli and $\partial$ is a derivative with respect to
a worldsheet coordinate. Furthermore we must consider $\lambda$ has an ultraviolet cut-off in the effective theory on
the vortex \cite{Shifman:2006kd}.

It is practically impossible to evaluate expression (\ref{kahler}), analytically or even numerically, in the general
case, but one can hope to do it in some particular simple examples. In this section we will derive the complete K\"ahler
potential for a single non-abelian semilocal vortex.

Using the set of coordinates defined by the moduli matrix formalism for a single non-Abelian semilocal vortex:
\beq H_0=\begin{pmatrix}
{\bf 1}_{\NC-1} & {\bf b} & 0 \cr 0 & z-z_0 & {\bf c}
\end{pmatrix}, \label{generalmoduli}
\eeq
we can determine the most general expression for the K\"ahler potential, compatible with the
$SU(\NC)_{\rm C+F}\times SU(\tNC)_{\rm F} \times U(1)$
isometry of the vacuum. Here $\NC-1$ column vector ${\bf b}$ and $\tNC$ row vector ${\bf c}$ are
moduli parameters. To this end we have to find the transformation properties of ${\bf b}$
and ${\bf c}$ under this symmetries. The moduli  matrix (\ref{generalmoduli}) transforms as:
\begin{eqnarray}
 \delta H_0&=&-H_0 \, u+v(u,z)\, H_0, \quad {\rm Tr}\, u={\rm Tr}\, v=0,\quad u^\dagger =-u\nn
u&=&\left(\begin{array}{ccc}
  \Lambda_{\NC-1} +i\,\lambda \, {\bf 1}_{N_{\rm C}-1}&  \bf{v}  & 0\\
-{\bf v}^\dagger  &-i(N_{\rm C}-1)\lambda &0\\
0&0&\tilde \Lambda_{\tNC}
       \end{array}\right),
\end{eqnarray}
where for simplicity $u$ is an infinitesimal $SU(\NC)_{\rm C+F}\times SU(\tNC)_{\rm F} \times U(1)$ transformation, and
$v(u,z)$ is an infinitesimal $V$-transformation that pulls back the matrix $H_0$ into the standard form of Eq.~(\ref{generalmoduli}). After
some calculations we find the following transformation properties for the moduli parameters:
\begin{eqnarray}
 \delta {\bf b}&=&
\Lambda_{\NC-1} \cdot {\bf b}+i \,\NC\, \lambda \, {\bf b}- {\bf v} - ( {\bf v} ^\dagger \cdot {\bf b}){\bf b}, \nn \delta
{\bf c} &=&-i(\NC-1)\, \lambda \,{\bf c} +({\bf v} ^\dagger \cdot {\bf b})\, {\bf c}-{\bf c} \cdot \tilde \Lambda_{\tNC} ,
\end{eqnarray}
from which one can infer:
    \begin{eqnarray}
    \delta \log\left(1+|{\bf b}|^2\right)=
    -({\bf v}^\dagger \cdot {\bf b})+{\rm c.c.},\ \
    \delta \log|{\bf c}|^2=(\bf{ v} ^\dagger \cdot {\bf  b})+{\rm c.c.},\ \
    \delta \left((1+|{\bf b}|^2)|{\bf c}|^2 \right) =0.
    \end{eqnarray}
These relations can be explained  if we note that $  ({c}_i, {c}_i {\bf b})$  (with arbitrary $i$, $i=\tNC $ for instance)  transforms
like a fundamental of $SU(\NC)_{\rm C+F}$  while   ${\bf c}$ as a fundamental of  $SU(\tNC)_{\rm F}$\footnote{One can verify this property using the transformation laws of  ${\bf b}$ and  ${\bf c}$. It is directly connected with the property of lump moduli spaces expressed by Eq.~(\ref{lumpsym})}.

\begin{figure}[ht]
\begin{tabular}{cc}
\includegraphics[width=7.7cm]{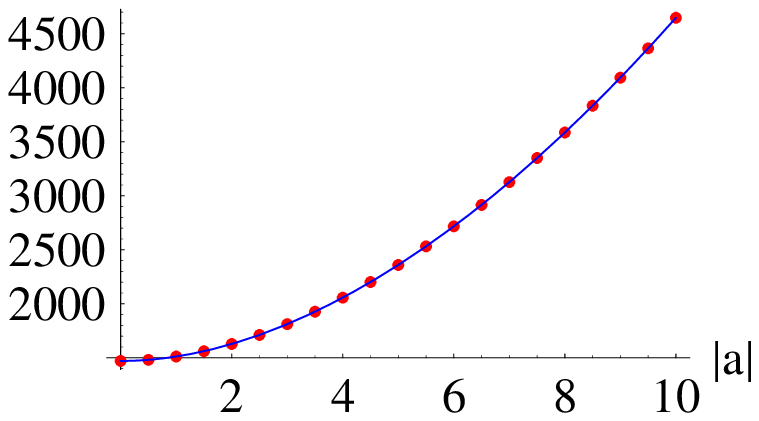} &
\includegraphics[width=7.7cm]{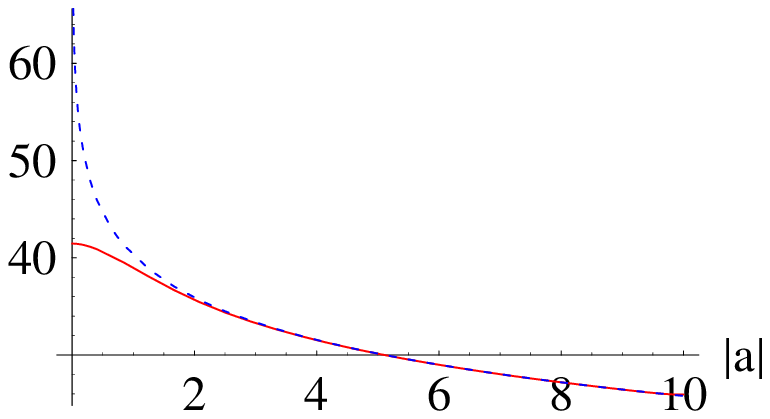}\\
{\small (a) K\"ahler potential} & {\small (b) K\"ahler metric }
\end{tabular}
\caption{{\small (a) The red dots are numerical computations of $F(|a|^2)$, while the blue line is an interpolation
done with the function shown in (\ref{interpolation}).} {\small (b) The red line is the K\"ahler metric $\partial_{a,
\bar a}K_{\rm abelian~semilocal}(z_0,a)$ for an abelian semilocal vortex at finite
 gauge coupling
\changed{
and is obtained from the interpolation function in (a),
while the blue dashed line is the
metric in the infinite gauge coupling limit (lump limit). The cut-off
 has been set to a very big value, $L=10^3$,
}
and
$g=\xi=1$.}}
\label{fig:Kahlerpic}
\end{figure}

Since the moduli parameters are zero modes related to the symmetry breaking of
$SU(\NC)_{\rm C+F}\times SU(\tNC)_{\rm F} \times U(1)$, the low energy action should be
invariant under the symmetry. In other words, the K\"ahler potential should be written in terms of
invariants under the transformation (up to K\"ahler transformation).
\changed{
The most general expression for the K\"ahler potential, up to K\"ahler
transformations, is thus given by
\beq
K(z_0,{\bf b}, {\bf c})=A|z_0|^2 + F (|a|^2)+B \log(1+|{\bf b}|^2),\label{generalkahler}
\eeq
where $A$ and $B$ are constants while $F(|a|^2)$ is an unknown function of the invariant
combination $|a|^2\equiv (1+|{\bf b}|^2)|{\bf c}|^2$. Note that a term $\log|{\bf c}|^2$ would also be invariant, but can be absorbed by a redefinition of
$F(|a|^2)$ and $B$. The constants and the function are determined as follows.
First of all, $z_0$ is the center of mass, so it is decoupled from any other modulus and its coefficient $A$ equals
half of the vortex mass,  $A = \pi \xi$.
Next let us consider the function $F(|a|^2)$.
Now,  if one fixes the orientational parameters to some
constant, e.g. ${\bf b}=0 \;(|a|^2=|{\bf c}|^2)$,  a non-abelian vortex becomes simply an embedding
of an abelian vortex into a larger gauge group, therefore the K\"ahler potential in (\ref{generalkahler}) must
reduce to  that of an abelian semilocal
vortex:
\beq
K(z_0,0, {\bf c})=K_{\rm abelian~semilocal}(z_0,{\bf c})=\pi \,\xi \, |z_0|^2+F(|{\bf c}|^2).
\eeq
It is important that the function $F(|a|^2)$ is independent of $\tNC(\ge 1)$, because the solution for $\tNC=1$ can
be embedded into those for $\tNC>1$. Furthermore, $F(|a|^2)$, written in term of the moduli parameters defined by the moduli matrix and defined as an integral over the configurations, should
  be smooth everywhere. 
In particular, in the limit  $|a|^2 \rightarrow 0$ it must be unique and equal just to
 that of the ANO vortex.  (A numerical result for $F(|a|^2)$ with $g=\xi=1$ and $L=10^3$ is shown in Fig.\ref{fig:Kahlerpic}). In this limit, which can be achieved letting  ${\bf c} \rightarrow 0$, }
 the vortex reduces to a local vortex, and also the K\"ahler potential should reduce
to that of a local vortex. $B$ is thus the K\"ahler class  of the non-Abelian vortex, $B = 4\pi/g^2$, as was found in
Ref.~\cite{Eto:2004rz}:
\beq
K(z_0,{\bf b}, 0)=K_{\rm non-abelian~local}(z_0,{\bf b})=\pi \xi |z_0|^2+\frac{4\pi}{g^2}\log(1+|{\bf b}|^2).
\eeq
This fixes the constants in (\ref{generalkahler}).
Therefore we find that the K\"ahler potential is determined uniquely in terms of that of an abelian semilocal
vortex:
\begin{eqnarray}
 K(z_0,{\bf b}, {\bf c}) & = & K_{\rm abelian~semilocal}(z_0,|a|)+\frac{4\pi}{g^2}\log(1+|{\bf b}|^2) \label{completekahler}
\end{eqnarray}
The function $F(|a|^2)$, being the K\"ahler potential for a single abelian vortex, can be computed numerically. Furthermore it is possible to find analytically the following behavior:
\begin{eqnarray}
 F(|a|^2)\sim\left\{
\begin{array}{cc}
\pi\,\xi \log\left(g^2\xi\,L^2 \alpha^{-1}\right)\times |a|^2 &\quad
{\rm for~} |a|\ll \frac{1}{g\sqrt{\xi}} \\
\pi\,  \xi \, |a|^2\left(\log\frac{L^2}{|a|^2}+1\right) +{\rm const.}& \quad {\rm for~} |a|\gg \frac{1}{g\sqrt{\xi}}.
\end{array}\right. ,
\end{eqnarray}
where $\alpha$ is some unknown  constant of ${\cal O}(1)$.
The behavior at small $a$ is simply a consequence of the smoothness of
the K\"ahler potential at $a=0$ and the cut-off dependence (\ref{divergent}).
Since the coefficient of the $|a|^2$ term is ${\cal O}(L^2)$, this term is dominant even for a medium region of $|a|$
as shown in Fig.\ref{fig:Kahlerpic}(a). The analytic form at large $a$ can be related to the expression of the
potential in the strong coupling limit as we will show in Eq.~(\ref{eq:geneKahler}) below.  It is very interesting to
note that we used a very simple function to interpolate the numerical results:
\beq
F(|a|)_{{\rm interp}}={\rm const}+ \pi\,  \xi \, \left(|a|^2+\frac{\alpha}{g^2
\xi}\right)\left(\log\frac{L^2}{|a|^2+\frac{\alpha}{g^2 \xi}}+1\right). \label{interpolation}
\eeq
This function gives a very good interpolation with only one relevant free parameter, $\alpha$, which appears to be of
order one. This interpolating function is nothing but the  K\"ahler potential in the lump limit
(Eq.~(\ref{eq:geneKahler})), regularized at $a=0$ by the introduction of a sort of UV cut-off:
$\rho_{eff}={\alpha}/{g^2 \xi}$. This appears to be a very nice, though empirical, way to show that local vortices act
as regularizers for small lump singularities.

Remarkably, in the case of a single vortex, the large-size limit is completely equivalent to the strong coupling limit.
To see this, note that the relevant quantity that triggers both limit is the following ratio:
\beq
R=\frac{\rho}{\rho_{\rm loc}},\quad \rho_{\rm loc}\equiv 1/(g\sqrt \xi),
\eeq
where $\rho$ is the physical size of the semilocal vortex, and $\rho_{\rm loc}$ is the typical size of a local vortex.
When the size moduli $\lambda$ vanish, the physical size $\rho$ shrink to $\rho_{\rm loc}$, so that $R\geq 1$. In the
strong coupling limit, $\rho_{\rm loc} \rightarrow 0$, and $R \rightarrow \infty$. The lump limit is thus really
defined by the limit $R \rightarrow \infty$, which can be achieved also at finite gauge coupling, just considering the
limit in which the physical size is very big: $\rho\sim\lambda \rightarrow \infty$. The lump K\"ahler potential (see
Eq.~(\ref{khalerlump})) is thus also a good approximation at finite gauge coupling, provided that we restrict to
solutions with a very big size. In fact one can see from Figure \ref{fig:Kahlerpic}(b) that this approximation is very
good also for small size\footnote{Presumably Eq.~(\ref{khalerlump}) should give a good approximation to the K\"ahler
potential also in the general case with several vortices, provided that we consider solutions with big typical sizes
(to this end we should not consider, for example, configurations such that $\det {H_0 H_0^{\dagger}}$ vanishes at some
vortex point).}.

In the lump limit the expression (\ref{kahler}) can be calculated analytically:
\beq
K \sim \xi {\rm Tr}\int\!\! d^2z \log\Omega =\xi \int\!\! d^2z \log ({\det {\Omega }})\sim \xi \int\!\! d^2z \log
({\det {H_0 H_0^{\dagger}} }).\label{khalerlump}
\eeq
Let us consider two particular, dual, examples: $k=1, \NC=2, \tNC=1\, (\NF=3)$. In this case all moduli are
non-normalizable \mnchanged{except for position moduli}, so we find no nontrivial dynamics on the vortex. Nonetheless
one may study the dynamics of these moduli by providing an infrared cut-off in (\ref{kahler}). Furthermore, according
to the above discussion, we can study the large size limit, in order to have an effective theory which can be
derived analytically. The expression (\ref{khalerlump}) is thus the K\"ahler potential for a $Gr_{2,3}\;(Gr_{1,3})$ lump
with topological charge $k=1$. In the abelian theory we find, from (\ref{matbelian}):
\beq
K_{\NC=1,\NF=3} = \xi \int_{|z|\le L} d^2z \log (|z-z_0|^2+|\tilde{b}|^2+|\tilde{c}|^2).
\eeq
\changed{
If we set $z_0=0$ for simplicity,
this integral is easily performed:
\beq
K_{\NC=1,\NF=3} = \xi \pi (|\tilde{b}|^2+|\tilde{c}|^2)\log \left( \frac {L^2}{|\tilde{b}|^2+|\tilde{c}|^2}
\right)+\xi \pi (|\tilde{b}|^2+|\tilde{c}|^2)
+{\cal O}(L^{-1}), \label{eq:k_nc1_nf3}
\eeq
where we omit divergent terms that do not depend on the moduli.
The corresponding metric is:
\beq
{L}_{\NC=1,\NF=3} = \xi \pi ( |\partial_\mu \tilde{b}|^2 + |\partial_\mu \tilde{c}|^2 ) \log {
\frac{L^2}{(|\tilde{b}|^2+|\tilde{c}|^2)}}+{\cal O}(L^0). \label{flat}
\eeq
}
 Note that we have obtained a conformally flat metric on ${\bf C}^2=\mathbb{R}^4$, which might be expected given the
$U(1)\times SU(2)_{F}$ isometry that acts on the parameters $\tilde{b}$ and $\tilde{c}$.

Now consider the non-abelian theory \mnchanged{with $\NC=2$}. The K\"ahler potential is given by:
\beq
K_{\NC=2,\NF=3}=\xi \pi |c|^2(1+|b|^2) \log{ \frac{L^2}{|c|^2(1+|b|^2)}}
+{\cal O}(L^0),
\label{eq:k_nc2_nf3}
\eeq
where we have used the moduli matrix coordinates defined in (\ref{moduli23}). Note that this potential is consistent
with the general expression we gave in (\ref{completekahler}), in the lump limit, up to logarithmic accuracy. The
$SU(2)_{C+F}\times U(1)$ symmetry of the theory, which leaves \nychanged{the quantity $|c|^2(1+|b|^2)$ invariant} is
again manifest. The metric that follows from this potential is:
\beq
\emph{L}_{\NC=2,\NF=3} =\xi \,\pi \, [ \, |c|^2 |\partial_\mu b|^2 + (1+|b|^2)|\partial_\mu c|^2 + (c\, b^{\dagger}\, \partial_\mu
c^{\dagger} \partial^\mu b+c. \, c.)] \, \log { \frac{L^2}{|c|^2(1+|b|^2)}}. \label{lump23}
\eeq
The expressions (\ref{flat}) and (\ref{lump23}) are related by the following change of coordinates:
\beq
\tilde{c}=c, \quad \tilde{b}=c \, b \quad (|c|^2(1+|b|^2)=|\tilde{c}|^2+|\tilde{b}|^2 \neq 0, \, c\neq0).
\eeq

The regularized metric of a semilocal vortex is  a conformally flat metric of ${\bf C}^2$ (modulo a change of
coordinates). This is valid, in the large size limit, for both dual theories: $\NC=2$, $\tNC=1$, $\NF=3$. This is
related to the fact that in both dual theories the semilocal vortex reduce to the same object, a
$Gr_{2,3}=Gr_{1,3}={\bf C}P^2$ lump.

The effective action in the lump limit, for generic $\NC$ and $\tNC$ can be found from the following K\"ahler
potential:
\beq
K_{\NC, N_{F}}=\xi \pi |{\bf c}|^2(1+|{\bf b}|^2) \log{ \frac{L^2}{|{\bf
c}|^2(1+|{\bf b}|^2)}}+\xi \pi\, |{\bf c}|^2(1+|{\bf b}|^2),
\label{eq:geneKahler}
\eeq
where ${\bf b}$ and ${\bf c}$ are vectors length $\NC-1$ and $\tNC$  respectively\footnote{Alternatively,
 vectors $\bf\tilde{b}$ and $\bf\tilde{c}$  of dimensions  $\tNC-1$ and $\NC$ can be used, as already emphasized.}.

\subsection{Duality and symmetry breaking}

In this subsection we make further comments on the effective actions we just obtained, with the aim of better illustrating  the meaning of the variables appearing in the effective actions, in terms of the symmetry breaking pattern due to the vortex
configuration.

For concreteness, we shall take the theory with  $\NC=1$ and $\NC=2$ with $\NF=3$.
In both sides of the dual,
 the vacuum ($k=0$) breaks  $G=SU(3)$ flavor symmetry to $H= SU(2) \times U(1)$.
The $SU(3)$ flavor symmetry in fact acts on the moduli
matrices for the vacuum configuration from  the right; $SU(2)\times U(1) \subset SU(3)$ can be absorbed by an appropriate
$V$-transformations acting on the left:
\beq
H_0^{\NC=1} = \left(1,\ 0,\ 0\right),\quad H_0^{\NC=2} = \left(
\begin{array}{ccc}
1 & 0 & 0\\
0 & 1 & 0
\end{array}
\right).
\eeq
The vacuum moduli spaces for both theories are the same complex manifold,  ${\bf C}P^2 \simeq G/H$.

Moduli matrices for $k=1$ vortex are given by  Eq.~(\ref{moduli23}) for $\NC=1$ and by  Eq.~(\ref{matbelian}) for
$\NC=3$. These moduli matrices  break the symmetry of the vacuum further. This spontaneous symmetry breaking
leads to the Nambu-Goldstone moduli, which are  the vortex orientation modes. In such a situation the worldsheet Lagrangian
of the $k=1$ vortex must be globally invariant under the symmetry of the vacuum,  $H = SU(2) \times U(1)$.
Furthermore the worldsheet theory must contain fields in definite  representations of $H$. Let us look at the transformation
law of the $\NC=1$ moduli parameters $\tilde b,\tilde c$ in Eq.~(\ref{matbelian}) and of the $\NC=2$ moduli parameters
$b,c$ in Eq.~(\ref{moduli23}).

In the abelian case the moduli matrix transforms in the following way:
\beq
H_0^{\NC=1}
\to \gamma^{\frac{2}{3}} (z,\ \tilde b,\ \tilde c) \left[ \left(
\begin{array}{ccc}
1 & & \\
& \alpha^* & \beta^*\\
& -\beta & \alpha
\end{array}
\right) \left(
\begin{array}{cc}
\gamma^{-\frac{2}{3}} & \\
&\gamma^{\frac{1}{3}} {\bf 1}_2
\end{array}
\right) \right],
\eeq
where $|\alpha|^2 + |\beta|^2=1$ and we have suppressed the uninteresting parameter $z_0$ corresponding to the position
of the vortex. Here the first factor $\gamma^{2/3}$ is an element of $V$ equivalence relation which is needed to keep
the coefficient of $z$ in $H_0$ equal to one, while the matrix product in the square bracket is an element of $SU(2)_F
\times U(1) \subset SU(3)$. The transformation properties of the moduli parameters are:
\beq
\tilde b \to \gamma \left(\alpha^* \tilde b -\beta \tilde c\right),\quad \tilde c \to \gamma \left(\beta^* \tilde b +
\alpha \tilde c\right).
\eeq
We see that $(\tilde b,\ \tilde c)$ transform linearly under $H$, and can be seen to form a $(\underline{2},1)$
representation of $SU(2)_F \times U(1)$. Notice that the $U(1)'$ subgroup in $SU(2)$,  defined by
\beq
\gamma' = \exp\left(i\frac{\lambda}{2}\vec{\tilde n}\cdot\vec\sigma\right) \quad\text{with}\quad \vec{\tilde n} =
\frac{1}{|\tilde b|^2+|\tilde c|^2}\left( \tilde b \tilde c^* + \tilde b^*\tilde c,\ i(\tilde b \tilde c^* - \tilde b^*
\tilde c),\ |\tilde b|^2 - |\tilde c|^2 \right),
\eeq
acts as $(\tilde b,\ \tilde c) \to e^{-i\frac{\lambda}{2}}(\tilde b,\ \tilde c)$, so that it can be always absorbed by
an overall $U(1)$ symmetry. The symmetry breaking given by the semilocal vortex is therefore
\beq
SU(2)_F\times U(1) \to U(1)'',
\eeq
where $U(1)''$ is the combination of $U(1)'$ and $U(1)$ keeping $(\tilde b,\tilde c)$ invariant.

On the other hand, the transformation law of the  moduli parameters $(b,c)$ in the $\NC=2$ moduli matrix
(\ref{moduli23}) is given by
\[
H_0^{\NC=2}
\to \left[ \gamma^{\frac{1}{3}} \left(
\begin{array}{cc}
\frac{1}{\alpha-b\beta} & 0\\
-\beta^*z & \alpha-b\beta
\end{array}
\right) \right] \left(
\begin{array}{ccc}
1 & -b & 0\\
0 & z & c
\end{array}
\right) \left[ \left(
\begin{array}{ccc}
\alpha & \beta &\\
-\beta^* & \alpha^* &\\
& & 1
\end{array}
\right) \left(
\begin{array}{cc}
\gamma^{\frac{-1}{3}}{\bf 1}_2 & \\
& \gamma^{\frac{2}{3}}
\end{array}
\right) \right],
\]
where the first factor is again an element of the $V$ equivalence needed to pull back the moduli matrix to its original
form. The last factor represents $SU(2)_{C+F}\times U(1) \subset SU(3)$. The transformation law of $(b,c)$ is thus:
\beq
b \to \frac{-\beta + \alpha^* b}{\alpha + \beta^* b},\qquad c \to (\alpha + \beta^* b)\gamma c.
\eeq
Since this is  highly non-linear, let us look carefully at its property. Let us first study the breaking pattern caused by
$b$. Since $b$ is invariant under $U(1)$, $SU(2)_{C+F}$ breaks to the $U(1)'$ subgroup defined by
\beq
\gamma' = \exp\left(i\frac{\lambda}{2}\vec n\cdot\vec\sigma\right) \quad\text{with}\quad \vec n =
\frac{1}{|b|^2+1}\left( b + b^*,\ i(b-b^*),\ |b|^2 - 1 \right). \label{coordinates}
\eeq
Thus the parameter $b$ describes  ${\bf C}P^1 \simeq SU(2)/U(1)'$ orientational moduli for $c=0$.
In the semilocal case $c\neq 0$, this $U(1)'$ symmetry is also broken.  $c$ is charged under
$U(1)'$ and transforms as $c \to e^{-i\frac{\lambda}{2}}c$. As this $U(1)'$  can be absorbed by $U(1)$ with $\gamma =
e^{i\frac{\lambda}{2}}$, the symmetry breaking pattern is actually
\beq
SU(2)_{C+F} \times U(1) \overset{b}{\longrightarrow} U(1)'\times U(1) \overset{c}{\longrightarrow} U(1)'',
\label{eq:sym_break}
\eeq
where $U(1)''$ is the combination of $U(1)$ and $U(1)'$ which leaves $c$ invariant.
The
topology of the moduli space for a single semilocal $U(2)$ vortex with $\NF=3$ thus is not a direct product $S^1 \times
S^2$, but a nontrivial fiber bundle $S^3 \sim S^1 \ltimes S^2$.

Summarizing, $(b,c)$ have a non-linear transformation law under the symmetries of the theory.
As long as  $c\neq0$ the coordinates change to the dual description:
\beq
\tilde b = bc,\quad \tilde c = c.
\eeq
can be made, which transform linearly as
%
$(\underline{2},1)$.   At the point  $c=0$ a change of coordinate
 (\ref{coordinates}) relates $b$ to   $\vec{n}$ which transforms as a triplet of $SU(2)$\footnote{An interpretation of the transformation of b
as a doublet under SU(2) is discussed in Ref.~\cite{Eto:2006dx}.}.
 This is an example  of the phenomenon in which, going from  local to semi-local vortex, the transformation properties of the fields appear to change.

 %

\subsection{Relation to the action of Shifman and Yung}

These discussions allow us to compare our result with that obtained earlier by Shifman and Yung in
\cite{Shifman:2006kd} more explicitly.  We  recall first  that the metric found by them is valid in the limit of large size, up to logarithmic terms.
This is precisely the same range of validity of the analytic results we found in the lump limit.
We found in that approximation a conformally flat metric on ${\bf C}^2$. Let us write this metric in terms of  the so-called Hopf coordinates:
\beq
\tilde{b}=\rho \, e^{i \xi_1} \sin \eta \quad \tilde{c}=\rho \, e^{i(\xi_1 - \xi_2)} \cos \eta, \quad (\rho \geq 0, \ 0 \leq \xi_1,\xi_2 \leq 2
\pi, \ 0\leq \eta \leq \pi/2).
\eeq
These coordinates describe $\mathbb{R}^4$ as $\mathbb{R} \times S^3$, and are useful to describe $S^3$ as the Hopf bundle $S^1\rightarrow S^3\rightarrow S^2$.
 In fact, the overall phase $\xi_1$ represents the $S^1$ fiber, while the following coordinates
\begin{eqnarray}
\nonumber n^1 &=& \rho^{-2} \Re (2 \, \tilde{b} \, \tilde{c}^{\dagger})= 2\, \sin \eta \cos \eta \cos \xi_{2}= \sin \theta \cos \phi; \\
 \nonumber n^2 &=& \rho^{-2} \Im (2 \,\tilde{b} \, \tilde{c}^{\dagger})=2 \, \sin \eta \cos \eta \sin \xi_{2}= \sin \theta \sin \phi; \\
 n^3 &=& \rho^{-2} (|\tilde{b}|^2-|\tilde{c}|^2)= (\sin^{2}\eta -\cos^2 \eta) = \cos \theta \label{sphere}
\end{eqnarray}
parameterize the $S^2$, where
\beq \theta \equiv \pi - 2 \,\eta, \qquad \phi \equiv \xi_{2} \eeq
can be identified with the usual spherical coordinates of the 2-sphere. The conformally flat
metric in the $(\rho,\xi_1,\phi,\theta)$ coordinates is:
\begin{eqnarray}
\nonumber \emph{L}_{\NC=2,\tNC=1} & = & \xi \pi \{ (\partial_\mu \rho)^2+\rho^2(\partial_\mu \xi_1)^2+
\frac{1}{4} \rho^2[(\partial_\mu \theta)^2+2(1 -\cos \theta)(\partial_\mu \phi)^2 ] \\
& - & \rho^2(1- \cos \theta) \partial_\mu \xi_1\partial^\mu \phi)
) \} \log { \frac{L^2}{\rho^2}}.\label{nonabmometric}
\end{eqnarray}
The relations with the non-abelian lump coordinates are:
\beq
b= e^{i \phi} \tan (\theta/2) \quad c=\rho \, e^{i(\xi_1 - \phi)} \cos (\theta/2).
\eeq
Note that the coordinate $b(\theta, \phi)$ parameterizes the $SU(2)_{C+F}/U(1)\sim S^2$ transformations of the
non-abelian orientation of the vortex. In the local case, this isometry is enough to generate the full target space of
a local vortex. This is why we find the metric of $S^2$ on these vortices:
\beq
L_{\NC=\NF=2}= \frac { \pi}{g^2} (\partial_\mu n^a)^2=\frac { \pi}{g^2} [\, (\partial_\mu \theta)^2+\sin^2
\theta(\partial_\mu \phi)^2\,].
\eeq

In the semilocal case the $SU(2)_{C+F}\sim S^3$ symmetry is completely broken. The moduli space of the semilocal vortex
contains an $S^3$ structure. Within this $S^3$, the $S^2$ of the $SU(2)_{C+F}$ orientation is combined in a nontrivial
way with a $U(1)$ phase. On the other hand, the authors of \cite{Shifman:2006kd} have found:
\beq
L_{SY}=\xi \pi \{ \frac {1}{4} |\rho|^2 (\partial_\mu n^a)^2 + | \partial_\mu \rho|^2 \} \log{ \frac {L^2}{|
\rho|^2}}, \label{shifm23}
\eeq
where now $\rho$ is a complex field.

Let us compare more explicitly the actions (\ref{shifm23}) and (\ref{nonabmometric}) near the point
$\vartheta\equiv\theta - \pi=0$, e.g., the point around which Shifman and Yung \cite{Shifman:2006kd} found an explicit
ansatz for the semilocal vortex. Eq.~(\ref{shifm23}) leads to
\beq
L_{SY} \simeq \xi \,\pi \, \left[ \frac {1}{4} |\rho|^2 [ \, (\partial_\mu \vartheta)^2+ \vartheta^2(\partial_\mu
\phi)^2 ] + |
\partial_\mu \rho|^2 \right] \log{ \frac {L^2}{|
\rho|^2}}\label{shifm23exp},
\eeq
while from (\ref{nonabmometric}) one finds
\beq
\nonumber L_{\NC=2,\tNC=1} \simeq \xi \, \pi \, \left[\,
\frac{1}{4} |\rho|^2 \, [\, (\partial_\mu \vartheta)^2+\vartheta^2(\partial_\mu \phi)^2 \,] + |\partial_\mu \rho|^2 -
  |\rho|^2 \frac {\vartheta^2}{2}\, \partial_\mu \,\xi_\rho \, \partial^\mu\, \phi \,\right] \, \log {
  \frac{L^2}{|\rho|^2}},\\\label{nonabetricexp}
\eeqq
where we have identified $\xi_1\equiv \xi_\rho$, the phase of the complex parameter $\rho$. Equations
(\ref{shifm23exp}) and (\ref{nonabetricexp}) look  very similar, and contain the same pieces of metric that describe
locally an $S^2$, but differ by a mixed term (the last term in the \mnchanged{square} brackets in
(\ref{nonabetricexp})), even at order $O(\vartheta^2)$.

Summarizing, the authors of \cite{Shifman:2006kd} assumed that there are no mixed kinetic terms
between the orientational moduli $n^a$ and the semilocal size $\rho$. The orientational moduli $n^a$ are obtained
implementing only $SU(2)_{C+F}$ rotations on their solution. Doing so they seem to have neglected the effects of these rotations on
$\rho$, which is nontrivial. Taking it into account should give rise to mixed terms, as in Eq.~(\ref{nonabmometric}).
In other words, those authors  appear to have found a metric on the trivialization $S^1 \times S^2$
of the bundle $S^3 \sim S^1 \ltimes S^2$.


\section{Summary and conclusions} \label{conclusion}

Generalizing Refs. \cite{Eto:2006pg}-\cite{Eto:2006cx} on the moduli space of the non-abelian vortices,
 we have analyzed the properties of {\it semilocal} vortices, appearing
in a $U(\NC)$ gauge theory with $\NF$ flavors of fundamental scalars, $\NF > \NC$,
in which the gauge group is completely broken in the presence of a Fayet-Iliopoulos term.
The moduli spaces of these semilocal vortices turn out to be (regularized)
holomorphic quotients, which are alternatively described as symplectic quotients upon symplectic reduction.
We have found a somewhat surprising and elegant relation between the moduli spaces of the semilocal vortices in Seiberg-like dual pair of theories, $U(\NC)$ and $U({\tNC})$: they correspond to two alternative regularizations of a ``parent'' space, which is not Hausdorff. In case of a fundamental (single) vortex the parent space is a weighted projective space with mixed weights. In the limit of lump ($g^{2}\to \infty$ or ${\tilde g}^{2} \to \infty$, respectively, in the two theories) these singular points become physically irrelevant, and the pair of dual moduli spaces degenerates into a common sigma model lump moduli space. As a byproduct we furnish a generalization of the rational map method to Grassmannian lumps.

We also studied the normalizable and non-normalizable zero-modes around these vortices (limiting ourselves to the bosonic modes) and discussed the low-energy effective actions associated to these degrees of freedom. In particular the relation between our result and that of Shifman and Yung \cite{Shifman:2006kd} has been clarified. Moreover, the precise number of normalizable moduli at a generic point of the moduli space has been provided, as well as an illustration of the mechanism responsible for its enhancement on special submanifolds.

These vortices were studied earlier in various papers \cite{Hanany:2003hp,Eto:2006pg,Eto:2006db,Tong:2005un,Shifman:2006kd} and our work is a natural extension. We hope that the new results obtained here will provide useful tools and hints for further
developments in the study of various topological solitons in non-abelian gauge theories and of their dynamics. In addition, we believe that our findings about the class of semilocal vortices can lead to the discovery of new appealing aspects of the close relationship between the dynamics of two-dimensional theories on the vortex world-sheet and the quantum dynamics of the underlying four-dimensional ${\cal N}=2$ gauge theory. The analysis of this correspondence is made possible by combining the knowledge of solitonic vortex strings and exact results coming from Seiberg-Witten curves, and has very recently received new impulse \cite{Eto:2006dx,Tong:2006pa,Ritz:2006rt,Edalati:2007vk,Tong:2007qj}.


\section*{Acknowledgments}
 We would like to thank M.~Shifman and A.~Yung for private communication leading to sensible improvement of Section~\ref{dyna}. We wish to thank K.~Hashimoto for fruitful discussions and collaboration. G.M.~and W.V.~want also to thank the Theoretical HEP
Group of the Tokyo Institute of Technology and the Theoretical Physics Laboratory of RIKEN for their warm hospitality.
G.M.~acknowledges the Foreign Graduate student Invitation Program (FGIP) of the Tokyo Institute of Technology.
W.V.~wishes to thank R.~Benedetti for precious discussions. Three of us (K.K., G.M., W.V.) wish to thank S.~B.~Gudnason
for discussions.  The work of K.O.~and N.Y.~is supported by the Research Fellowships of the Japan Society for the
Promotion of Science for Young Scientists. The work of M.E.~is also supported by the Research Fellowships of the Japan
Society for the Promotion of Science for Research Abroad.

\appendix

\section{Weighted projective spaces with mixed weights} \label{wpsnw}

In this appendix we provide a pedagogical introduction to weighted projective spaces with both positive and negative weights, starting from the simplest example and then stepping up the complexity.

\subsection{$W{\bf C}P^1_{(1,-1)}$} \label{wcp1-1}

First of all consider the simplest example, $W{\bf C}P^1_{(1,-1)}$. Let us try to define this space
 na\"{i}vely as: \beq W{\bf
C}P^1_{(1,-1)}=\{(y_1,y_2)\sim(\lambda y_1,\lambda ^{-1} y_2),\quad \lambda \ne 0, \quad
(y_1,y_2)\neq(0,0) \},
\eeq
in analogy with a weighted projective
space with positive integer weights.
Apparently we need two patches:
 \beq (1,a), \quad y_1 \neq 0, \quad
a=y_1 y_2; \\
(b,1), \quad y_2 \neq 0, \quad b=y_1 y_2. \eeq
But as the transition function is trivial
\beq a=b, \eeq
 the points $(1,a)$ and $(a,1)$ are actually the \emph{same} point and one is tempted to conclude that
 this space is simply equivalent to ${\bf C}$. This is not quite so. Two points
 $(1,a)$ and $(a,1)$ are the same point only for $a \neq0$.
The two points $(1,0)$ and $(0,1)$ are {\it distinct} points, having nevertheless no disjoint open neighborhoods,
making the space non-Hausdorff. In other words, our na\"{i}ve $W{\bf C}P^1_{(1,-1)}$ is ${\bf C}$ plus a point, $\{(1,a)\}\cup
\{(0,1)\}$.
A good remedy is simply to eliminate ``by hand'' one
of the points in the definition of the space. Note that now only one patch suffices to cover the whole space, and one finds
\beq W{\bf C}P^1_{(\underline{1},-1)}=\{(y_1,y_2)\sim(\lambda y_1,\lambda ^{-1} y_2),\;
y_1\neq 0 \}={\bf C}.
\eeq
By replacing the condition $y_1\neq 0$ by $y_2\neq 0$ one gets the space $W{\bf
C}P^1_{(1,\underline{-1})}$, which is still be isomorphic to ${\bf C}$. In the above we introduced a notation indicating the
underlined coordinates are those which cannot all vanish.

This space turns out in fact to be the moduli space of an abelian semilocal vortex with two
flavors:
\beq
{\cal M}_{k=1,N=1,\NF=2}=W{\bf C}P^1_{(\underline{1},-1)}={\bf C}.
\eeq

\subsection{$W{\bf C}P^2_{(1,1,-1)}$}

Consider now a more interesting case,
\beq W{\bf C}P^2_{(1,1,-1)}=\{(y_1,y_2,y_3)\sim(\lambda y_1,\lambda y_2, \lambda ^{-1} y_2),\quad \lambda \ne 0, \quad
(y_1,y_2, y_{3}) \neq(0,0,0) \}.
\eeq

   In particular, consider a point $(a \,\epsilon, b \, \epsilon,1)$. By taking $\epsilon$ arbitrarily small,
this point can be made arbitrarily close to $(0,0,1)$. But since
\beq (a \,\epsilon, b\, \epsilon,1)\sim
(a,b,\epsilon), \eeq
this point is arbitrarily close to a point $(a,b,0)$ in the subspace ${\bf C}P^1$ subspace also.
In order to make the space Hausdorff,
one must either
eliminate the point $(0,0,1)$ or extinguish the whole ${\bf C}P^1$ made of the points $(a,b,0)$.
Eliminating the entire ${\bf C}P^1$, one needs only one patch to describe the entire space:
\beq
W{\bf C}P^2_{(1,1,\underline{-1})}=(a,b,1)={\bf C}^2, \eeq
{\it i.e.}, just the two dimensional complex space.

 The first possibility is more interesting. One takes
\beq W{\bf C}P^2_{(\underline{1,1},-1)}=\{(y_1,y_2,y_3)\sim(\lambda y_1,\lambda y_2, \lambda ^{-1} y_2),\quad
(y_1,y_2)\neq(0,0) \}.
\eeq
Now one needs two patches to cover the whole space:
\beq (1,a,b),\quad y_1\neq
0, \quad a=y_2/y_1,\quad b=y_1 y_3,\\
(a',1,b'),\quad y_2\neq 0,\quad a'=y_1/y_2,\quad b'=y_2 y_3.
\eeq
The transition functions are
\beq a=\frac{1}{a'},\quad b=a' b'. \eeq
  To understand better the nature of this space, imagine that we
add again the point $(0,0,1)$ and eliminate the ${\bf C}P^1=(y_1,y_2,0)$: we obtain ${\bf C}^2$. This is something like the
inverse procedure of blowing-up ${\bf C}^2$, namely a blow-down. In fact the blow-up of ${\bf C}^2$ is obtained substituting a point in ${\bf C}^2$ with a sphere
 ${\bf C}P^1$.

\subsection{The blow-up of ${\bf C}^2$: $\tilde{{\bf C}}^2$}
\label{blow up}

The blow-up of ${\bf C}^2$ is given in terms of a projection map
$\Gamma:{\bf C}^2\times{\bf C}P^1\rightarrow {\bf C}^2$ defined as
the points that satisfy
\beq \label{blow up c2} \Gamma : \{(x_1,x_2,y_1,y_2) \, | \,x_1 y_2=x_2 y_1 \},
\eeq
where $(x_1,x_2)$ are the coordinates of ${\bf C}^2$ and $(y_1,y_2)$ are the homogeneous coordinates of ${\bf C}P^1$. Consider
first the patch:
\beq (x_1,x_2,1,a),\quad a=y_2/y_1,
\eeq in which we solve the
constraint $(x_1,x_2,y_1,y_2)$ as
\beq b=x_1,\quad x_2=a b. \eeq
 In
the other patch we have:
\beq (x_1,x_2,a',1),\quad a'=y_1/y_2 \eeq
with
\beq b'=x_2,\quad x_1=b' a'. \eeq
Now the transition functions
are readily found:
\beq a=1/a',\quad b=a' b'. \eeq
We see that
transition functions of the blow-up $\tilde{{\bf C}}^2$ and that of
the \nychanged{$W{\bf C}P^2_{(\underline{1,1},-1)}$} coincide exactly.

Thus we conclude:
 \beq {\cal M}_{\NC=2,\NF=3;k=1}=W{\bf C}P^2_{(\underline{1,1},-1)}=\tilde{{\bf C}}^2.
\eeq

\subsection{$W{\bf C}P^n_{(1,1,\dots,1,-1)}$}
The generalization for a generic weighted projective space with one
negative weight is straightforward. It turns out that the
non-abelian semilocal vortices with $k=1$ and $\NF=\NC+1$
have moduli given by:
\beq \nychanged{{\cal M}_{\NC,\NF=\NC+1;k=1}=W{\bf
C}P^{\NC}_{(\underline{1,1,\dots,1},{-1})}=\tilde{{\bf C}}^{\NC}.}
\eeq

\subsection{$W{\bf C}P^n_{(1,1,\dots,1,-1,\dots,-1)}$} \label{rescon}

Consider the simplest case \nychanged{among general cases with
both the multiple positive and negative weights}:
$W{\bf C}P^3_{(\underline{1,1},-1,-1)}$. This is
defined as ${\bf C}^4\setminus \{(0,0,v,w)\}$ modulo the equivalence
$$
(y_1,y_2,y_3,y_4) \sim (\lam y_1,\lam y_2,\lam^{-1} y_3,\lam^{-1}
y_4).
$$
Only two patches are needed to cover the whole space:
\beq
(1,a,b,c),\quad y_1\neq 0, \\
(a^{\prime},1,b^{\prime},c^{\prime}),\quad y_2\neq 0. \eeq
The transition functions are
\beq a={1 / a^{\prime}}, \quad
b=b^{\prime}a^{\prime},\quad c=c^{\prime}a^{\prime}. \eeq
One may represent this variety as follows
  \beq W{\bf C}P^3_{(1,1,-1,-1)}=\{ {\bf
C}^2(x_1,x_2)\times {\bf C}^2(x_3,x_4) \times {\bf C}P^1(y_1,y_2) / \; x_1 y_2=x_2 y_1,\; x_3 y_2=x_4 y_1 \}.\eeq
 One can check that
this is correct following the same procedure as in Appendix~\ref{blow up}. The above set is nothing but the resolved conifold (well known to physicists since Ref.~\cite{Candelas:1989js} and in the context of AdS/CFT \cite{PandoZayas:2000sq}). In fact, the
equation
\beq x_1x_4=x_2 x_3 \eeq
 always holds. When the $x_i$ are not all zero and fixed, a point in ${\bf C}P^1$ is fixed. At
the origin ($x_i=0 \, \forall i$), instead, the singularity of the conifold is replaced by the full ${\bf C}P^1$. What happens is
the following: the conifold is topologically a cone over $S^2\times S^3$ and both spheres degenerate at the tip of the cone; by blowing up the $S^2$ we
get the resolved conifold (on the contrary, replacing the singularity with an $S^3$ leads to the deformed conifold, see the
Klebanov-Strassler solution \cite{Klebanov:2000hb}). This case corresponds to the $U(2)$ non-abelian semilocal vortex with \nychanged{four flavors}:
\beq {\cal M}_{\NC=2,\NF=4;k=1}=W{\bf C}P^3_{(1,1,-1,-1)}=\hbox{resolved conifold}. \eeq


\subsection{$W{\bf C}P^n_{(2,1,\dots,1,-1)}$}

To study this case we must remember what we have already learned
about weight 2 and about negative weight. The presence of a weight 2
leads to a $Z_2$ symmetry and, as a consequence, a conical
singularity. The following statement is straightforward:
\beq W{\bf
C}P^n_{(2,1,\dots,1,-1)}=W{\bf C}P^n_{(1,1,\dots,1,-1)}/\mathbb{Z}_2
\eeq
with an obvious $\mathbb{Z}_2$ action. In fact, combining this
relation with the results of the previous sections we find:
\beq
W{\bf C}P^n_{(2,1,\dots,1,-1)}=\tilde{{\bf C}}^n/ \mathbb{Z}_2. \eeq
To determine the $\mathbb{Z}_2$ action on the second term we can
simply identify coordinates of the two spaces. Consider for example
$W{\bf C}P^2_{(2,1,-1)}$. This is equivalent to $W{\bf C}P^2_{(1,1,-1)}/
\mathbb{Z}_2$, where the discrete symmetry identifies $W{\bf
C}P^2_{(1,1,-1)}(\pm y_1,y_2,y_3,y_4)$. We can write down the
inhomogeneous coordinates and readily find the $\mathbb{Z}_2$
action on this coordinates.
\beq (1,a,b),\quad y_1\neq 0,
\quad a=\pm y_2/y_1,\quad b=\pm y_1 y_3, \\
(a',1,b'),\quad y_2\neq 0,\quad a'=\pm y_1/y_2,\quad b'=y_2 y_3. \eeq
Now remember the definitions of the blown-up $\tilde{{\bf C}}^2$
coordinates in terms of that of ${\bf C}^2(x_1,x_2)$:
\beq
b=x_1,\quad x_2=a b, \\
b'=x_2,\quad x_1=a' b'.
\eeq
From this last relations we see that we must consider a
$\mathbb{Z}_2$ action on $x_1$ (not on $x_2$!). Thus we have:
 \beq
\tilde{{\bf C}}^2 / \mathbb{Z}_2 : \{(\pm x_1,x_2,\pm y_1,y_2) \, |
\,x_1 y_2=x_2 y_1 \}. \eeq In fact we would like to blow-up ${\bf
C}^2$ by substituting the origin with a $W{\bf C}P^{1}_{(2,1)}$. To do
this we are forced to consider a $\mathbb{Z}_2$ on ${\bf C}^2$ as well.

We can also directly consider $W{\bf C}P^2_{(2,1,-1)}$, \nychanged{whose}
transition functions are:
\beq (1,a,b),\quad y_1\neq 0, \quad a=\pm
y_2/\sqrt{y_1},\quad b=\pm \sqrt {y_1} y_3,
\\ (a',1,b'),\quad y_2\neq 0,\quad a'=y_1/y_2^2,\quad b'=y_2 y_3, \\
a'=1/a^2 \; (a=\pm 1/\sqrt{a'}), \quad b'=a b \; (b=\pm \sqrt{a'} b').
\eeq Now if we want to consider this space as a blow up of ${\bf
C}^2$ we consider a projection $\Gamma:{\bf C}^2\times W{\bf
C}P^1_{(2,1)}\rightarrow {\bf C}^2$ and remembering that $y_1$ has
weight 2:
\beq \Gamma :
\{(x_1,x_2,y_1,y_2) \, | \,x_1 y_2= \pm x_2 \sqrt{y_1} \}. \eeq
Again we have two patches:
\beq (x_1,x_2,1,a),\quad a=\pm y_2/\sqrt {y_1}, \; b=x_1, \; x_2= a b
\eeq
and
\beq (x_1,x_2,a',1),\quad a'= y_1/y_2^2, \; b'=x_2, \;
x_1=\pm \sqrt {a'} b'. \eeq The relation $x_1 y_2= x_2 \sqrt{y_1}$
is consistent with $(x_1=b,a)= (-x_1=-b,-a).$

This consistent
relations leads to the correct transition functions. In the end we have:
\beq \Gamma:{\bf C}^2/\mathbb{Z}_2\times \nychanged{W{\bf C}P^1_{(2,1)}}
\rightarrow \tilde{{\bf C}}^2/\mathbb{Z}_2.\eeq
In general the action of the discrete symmetries should be
simultaneous on both ${\bf C}^n$ and $W{\bf C}P^{n-1}$. Let us
consider in detail the space $W{\bf C}P^3_{(2,1,1,-1)}$, that is
important in the \nychanged{$U(2)$} case with
one additional flavor \nychanged{(see Appendix~\ref{compsemvort})}.
We can consider this space as the blow-up $\tilde{{\bf
C}}^3/ \mathbb{Z}_2$. Following the considerations of the previous
section we can write down the transitions functions \nychanged{for}
$W{\bf C}P^3_{(2,1,1,-1)}\sim(y_1,y_2,y_3,y_4)$:
\nychanged{
\beq (1,X,Y,Z),\quad
y_1\neq 0, \quad X=\pm y_2/\sqrt{y_1},\quad Y=\pm
y_3/\sqrt{y_1},\quad Z=\pm \sqrt {y_1} y_4,
\\ (a,1,b,c),\quad y_2\neq 0,\quad a=y_1/y_2^2,\quad b=y_3/y_2,
\quad c=y_2 y_4, \\
(a',b',1,c),\quad y_3\neq 0,\quad a'=y_1/y_3^2,\quad b'=y_2/y_3,\quad
c'=y_3 y_4, \\
X=\pm 1/\sqrt{a}, \quad Y= \pm b / \sqrt{a},\quad Z= \pm
\sqrt{a} b \quad\textrm{and}\quad a'=a/b^2,\quad b'=1/b, \quad c'=c b.
\eeq
}
The remaining transition function can be
obtained from the relations above. To construct the
blow-up of ${\bf C}^3/ \mathbb{Z}_2$ with our $W{\bf
C}P^2_{(2,1,1)}$ we define: \beq \Gamma :
\{(x_1,x_2,x_3,y_1,y_2,y_3) \, | \,x_1 y_2= \pm x_2 \sqrt{y_1}, \;
x_1 y_3= \pm x_3 \sqrt {y_1}, (x_2 y_3=x_3 y_2) \}. \eeq The action
of $\mathbb{Z}_2$ on ${\bf C}^3$ is found as in the previous case:

\beq \tilde{{\bf C}}^3/ \mathbb{Z}_2 : \{(\pm x_1,x_2,x_3,\pm
y_1,y_2,y_3) \, | \,x_1 y_2= x_2 y_1, \; x_1 y_3= x_3 y_1, (x_2
y_3=x_3 y_2) \}. \eeq This space has two fixed submanifold, an
entire $\tilde{{\bf C}}^2$, when $x_1=y_1=0$ and the point
$x_1=x_2=x_3=y_2=y_3=0$.


\section{Composing semilocal vortices}\label{compsemvort}
In this appendix we consider in detail the case of $k=2$ vortices, with $\NC=2$ and $\tNC=1$. We find explicitly the corresponding
moduli matrix and the matrices ${\bf Z}$, ${\bf \Psi}$ and ${\bf \tilde{\Psi}}$. Then we consider the case of coincident
vortices, studying in detail the properties of the resulting moduli space.

\subsection{Moduli Matrix and K\"ahler quotient construction }
\begin{itemize}
\item \underline{$\NC=2, \NF \ge 3$}

The moduli space is described by three $2 \times \NF$ moduli matrices, one for each patch needed to describe the whole space. The
constraint that $H_0$ must satisfy is:
\beq
\det{H_0 H_0^{\dagger}}\sim |z|^4, \quad |z| \rightarrow \infty.
\eeq
Taking into account the fact that we can fix the $V(z)$ equivalence:
\beq
H_0(z) \to V(z) \, H_0(z),
\eeq
we can put $H_0$ into an upper triangular form. The most general moduli matrix is:
\beq
H_0^{(0,2)}(z) &=& \left(
\begin{array}{c|c}
{\bf D}^{(0,2)}(z) & {\bf Q}^{(0,2)}(z)
\end{array}
\right) = \left(
\begin{array}{cc|c}
1 & -a z - b & - a {\bf q} \\
0 & z^2- \alpha z-\beta & {\bf q} z + {\bf p}
\end{array}
\right),\label{20patch}\\
H_0^{(1,1)}(z) &=& \left(
\begin{array}{c|c}
{\bf D}^{(1,1)}(z) & {\bf Q}^{(1,1)}(z)
\end{array}
\right) = \left(
\begin{array}{cc|c}
z -\phi & -\eta & {\bf s} \\
-\tilde {\eta} & z-\tilde {\phi} & {\bf t}
\end{array}
\right),\label{11patch}\\
H_0^{(2,0)}(z) &=& \left(
\begin{array}{c|c}
{\bf D}^{(2,0)}(z) & {\bf Q}^{(2,0)}(z)
\end{array}
\right) = \left(
\begin{array}{cc|c}
z^2-\alpha' z-\beta' & 0 & {\bf q}'z+{\bf p}' \\
-a'z - b' & 1 & -a'{\bf q}'
\end{array}
\right).\label{02patch}
\eeq
Here all of $\{{\bf q},{\bf p},{\bf s},{\mathbf{t}},{\bf q}',{\bf p}'\}$ are row $(\NF-2)$-vectors.

From the moduli matrix written above we can extract the three matrices ${\bf Z}_{[2\times2]}$, ${\bf \Psi}_{[2 \times 2]}$ and
${\bf \tilde{\Psi}}_{[2 \times (\NF-2)]}$, defined modulo the equivalence relation:
\beq
\left({\bf Z}, {\bf \Psi}, \tilde{\bf \Psi}\right) \sim \left({\cal V}{\bf Z}{\cal V}^{-1}, {\bf \Psi}{\cal V}^{-1}, {\cal
V}\tilde{\bf \Psi}\right), \qquad {\cal V} \in GL(2,{\bf C}).
\eeq
Following the scheme sketched in Section~\ref{setup}, from:
\beq
{\bf D}(z) {\bf\Phi}(z) = {\bf J}(z) P(z) =0 \mod P(z) \label{eigenmatrix1bis}
\eeq
we find the matrices ${\bf\Phi}(z)$ and ${\bf J}(z)$:
\beq
{\bf \Phi}^{(0,2)}(z) = \left(
\begin{array}{cc}
b z-b \alpha+a \beta & az + b\\
z-\alpha & 1
\end{array}
\right), \quad {\bf J}^{(0,2)}(z) = \left(
\begin{array}{cc}
-a & 0 \\
z- \alpha & 1
\end{array}
\right);
\eeq
\beq
{\bf \Phi}^{(1,1)}(z) = \left(
\begin{array}{cc}
z - \tilde \phi & \eta \\
\tilde \eta & z + \phi
\end{array}
\right), \quad {\bf J}^{(1,1)}(z) = \left(
\begin{array}{cc}
1 & 0 \\
0 & 1
\end{array}
\right);
\eeq
\beq
{\bf \Phi}^{(2,0)}(z) = \left(
\begin{array}{cc}
z-\alpha' & 1 \\
b'z-b' \alpha' & a'z + b'
\end{array}
\right), \quad {\bf J}^{(2,0)}(z) = \left(
\begin{array}{cc}
z-\alpha' & 1 \\
-a' & 0
\end{array}
\right);
\eeq
Now we use the matrices ${\bf \Phi}(z)$ and ${\bf J}(z)$ to obtain ${\bf Z}$ and ${\bf \Psi}$ from the following relation:
\beq
z {\bf \Phi}(z) = {\bf \Phi}(z) {\bf Z}+ P(z){\bf \Psi}. \label{eigenmatrix3} \eeq We have:
\beq
{\bf Z}^{(0,2)} = \left(
\begin{array}{cc}
0 & 1\\
\beta & \alpha
\end{array}
\right),\quad {\bf \Psi}^{(0,2)} = \left(
\begin{array}{cc}
b & a\\
1 & 0
\end{array}
\right);
\eeq

\beq
{\bf Z}^{(1,1)} = \left(
\begin{array}{cc}
\phi & \eta\\
\tilde \eta & \tilde \phi
\end{array}
\right),\quad {\bf \Psi}^{(1,1)} = \left(
\begin{array}{cc}
1 & 0 \\
0 & 1
\end{array}
\right);
\eeq
\beq
{\bf Z}^{(2,0)} = \left(
\begin{array}{cc}
0 & 1\\
\beta' & \alpha'
\end{array}
\right),\quad {\bf \Psi}^{(2,0)} = \left(
\begin{array}{cc}
1 & 0 \\
b' & a'
\end{array}
\right).
\eeq

The additional semilocal moduli, contained in the matrices ${\bf Q}^{(0,2)}$, can be extracted by:
\beq
{\bf Q}(z) = {\bf J}(z) \tilde {\bf \Psi} \label{eigenmatrix4}\eeq In fact we find
\beq
\tilde{\bf \Psi}^{(0,2)} = \left(
\begin{array}{c}
{\bf q}\\
\alpha {\bf q} + {\bf p}
\end{array}
\right) \quad \tilde{\bf \Psi}^{(1,1)} = \left(
\begin{array}{c}
{\bf s} \\
{\bf t}
\end{array}
\right) \quad \tilde {\bf \Psi}^{(2,0)} = \left(
\begin{array}{c}
{\bf q}'\\
\alpha'{\bf q}' + {\bf p}'
\end{array}
\right).
\eeq

\item \underline{$\NC=1, \NF=3$}

The moduli space is given by a $1 \times 3$ moduli matrix that satisfy the same boundary conditions as in the previous case. The
most general matrix of this kind is:
\beq
H_0^{(0,2)}(z) &=& \left(
\begin{array}{c|c}
{\bf D}^{(0,2)}(z) & {\bf Q}^{(0,2)}(z)
\end{array}
\right) =\left(
           \begin{array}{c|cc}
             z^2- \alpha z-\beta & q_1 z+p_1 & q_2 z+p_2 \\
           \end{array}
         \right).
\eeq
From (\ref{eigenmatrix1bis}) we easily get:
\beq
{ \bf \Phi}(z)= { \bf J}(z) =\left(
                                \begin{array}{cc}
                                  z & 1 \\
                                \end{array}
                              \right),
\eeq
and from (\ref{eigenmatrix3}):
\beq
{\bf Z}=\left(
          \begin{array}{cc}
            \alpha & 1 \\
            \beta & 0 \\
          \end{array}
        \right),\quad {\bf \Psi}=\left(
                        \begin{array}{cc}
                          1 & 0 \\
                        \end{array}
                      \right).
\eeq
Finally from (\ref{eigenmatrix4}):
\beq
\tilde{\bf \Psi}=\left(
                   \begin{array}{cc}
                     q_1 & q_2 \\
                     p_1 & p_2 \\
                   \end{array}
                 \right)
\eeq
\end{itemize}

\subsection{Coincident (axially symmetric) semilocal vortices}

Here we explore in detail the case of coincident semilocal vortices, generalizing the approach of
\cite{Eto:2006cx}. We will focus on the case $k=\NC=2, \ \NF=3$, being the generalization to an arbitrary number of color and
flavor straightforward.

 In the case of coincident vortices we can write for the matrix ${\bf Z}$:
\beq
{\bf Z} = \epsilon vv^T, \quad \epsilon =\left(
                                          \begin{array}{cc}
                                            0 & 1 \\
                                            -1 & 0 \\
                                          \end{array}
                                        \right)
\eeq
so that $v$ transforms as a fundamental vector:
\beq
\epsilon vv^T \sim {\cal V} \epsilon vv^T {\cal V}^{-1} = \epsilon ({\cal V}^{-1})^T vv^T {\cal V}^{-1} \quad \rightarrow \quad v
\sim \left(\det {\cal V}\right)^{\frac{1}{2}}({\cal V}^{-1})^T v
\eeq
where we used $\epsilon {\cal V} = \det {\cal V} \times ({\cal V}^{-1})^T \epsilon$ for ${\cal V} \in GL(2,{\bf C})$. Rewrite
$\lambda {\cal S} = ({\cal V}^{-1})^T$, then we get
\beq
M = \left( {\bf \Psi}^T,\ v,\ \epsilon \tilde {\bf \Psi} \right) \sim {\cal S}\left( \lambda {\bf \Psi}^T,\ v,\
\lambda^{-1}\epsilon \tilde {\bf \Psi} \right), \label{semilocalset}\eeq where we have used
\beq
\epsilon \tilde {\bf \Psi} \to \epsilon {\cal V} \tilde {\bf \Psi} = \det {\cal V} \times ({\cal V}^{-1})^T\epsilon \tilde {\bf
\Psi} = \frac{1}{\lambda} {\cal S} \epsilon \tilde{\bf \Psi}.
\eeq
Thus, the set \nychanged{(\ref{semilocalset})} gives a weighted Grassmannian manifold with negative weights:
\beq
\tilde{\cal M}_{\NC,\tilde\NC,k=2} = WGr_{\NC + \tilde\NC + 1,2}^{(1\times \NC,0,-1\times \tilde\NC)} = WGr_{\NF + 1,2}^{(1\times
\NC,0,-1\times \tilde\NC)}.
\eeq
\begin{itemize}
\item \underline{$\NC=2, \NF=3$}

The results of the previous section, in the case of coincident vortices, can be collected in the following matrices:
\beq
&& M^{(0,2)} = \left(
\begin{array}{cccc}
b & 1 & 0 & p \\
a & 0 & 1 & -q
\end{array}
\right),\ M^{(1,1)} = \left(
\begin{array}{cccc}
1 & 0 & -Y & \xi\\
0 & 1 & X & -\eta
\end{array}
\right),\ \nn && M^{(2,0)} = \left(
\begin{array}{cccc}
1 & b' & 0 & p'\\
0 & a' & 1 & -q'
\end{array}
\right).
\eeq
These can be considered as the three patches which describe the "regularized" weighted Grassmannian
$WGr_{4,2}^{(\underline{1,1},0,-1)}$. Notice that there exists a ${\bf Z}_2$ symmetry in the patch $M^{(1,1)}$: $(X,Y) \to -
(X,Y)$.

More natural coordinates on this manifold are given by the Pl\"ucker coordinates
\beq
&&\left(
\begin{array}{c}
d_{12}\\
d_{23}\\
d_{13}\\
d_{14}\\
d_{24}\\
d_{34}
\end{array}
\right) \equiv \left(
\begin{array}{c}
\det M_{[12]}\\
\det M_{[23]}\\
\det M_{[13]}\\
\det M_{[14]}\\
\det M_{[24]}\\
\det M_{[34]}
\end{array}
\right) \sim \left(
\begin{array}{c}
\lambda^2\det M_{[12]}\\
\lambda\det M_{[23]}\\
\lambda\det M_{[13]}\\
\det M_{[14]}\\
\det M_{[24]}\\
\lambda^{-1}\det M_{[34]}
\end{array}
\right)\nonumber\\
&&\sim \left(
\begin{array}{c}
-a\\
1\\
b\\
-b q - ap\\
-q\\
-p
\end{array}
\right) \sim \left(
\begin{array}{c}
1\\
Y\\
X\\
-\eta\\
-\xi\\
Y\eta - X\xi
\end{array}
\right) \sim \left(
\begin{array}{c}
a'\\
b'\\
1\\
-q'\\
-b'q' - a'p'\\
-p'
\end{array}
\right), \label{grassmcollect}
\eeq in which we used the Pl\"ucker identity:
\beq
d_{12}d_{34} - d_{13}d_{24} + d_{14}d_{23} = 0 \label{pluecker}
\eeq

The transition functions can be easily read from this. For example, those from $M^{(1,1)}$ to $M^{(2,0)}$ are:
\beq
a = - \frac{1}{Y^2},\quad b = \frac{X}{Y},\quad q = \xi,\quad p = XY\xi - Y^2\eta,\quad \left(bq + qp = \eta\right).
\eeq
Similarly, we can easily find transition functions from $M^{(0,2)}$ to $M^{(2,0)}$
\beq
a = -\frac{a'}{{b'}^2},\quad b = \frac{1}{b'},\quad q = b'q' + a'p',\quad p = b'p',\quad \left(bq+ap = q'\right).
\eeq

It is known that we can consider a \nychanged{Grassmannian} manifold as an embedding into a bigger projective space defined by the Pl\"ucker
coordinates themselves. In our case we have the following equivalence relation:
\beq &&
\begin{array}{c}
\left( d_{12},\ d_{23},\ d_{13},\ d_{14},\ d_{24},\ d_{34} \right)
\end{array}
\nonumber\\
&&\sim
\begin{array}{c}
\left( \lambda^2 d_{12},\ \lambda d_{23},\ \lambda d_{13},\ d_{14},\ d_{24},\ \lambda^{-1} d_{34} \right)
\end{array}
\eeq
That is a $W{\bf C}P^5_{(2,1,1,0,0,-1)}$. There are two coordinates with 0 weight. These two coordinates are in fact
inhomogeneous. We can therefore write:
\beq W{\bf C}P^5_{(2,1,1,0,0,-1)}=W{\bf C}P^3_{(2,1,1,-1)}(d_{12},d_{13},d_{23},d_{34})
\times {\bf C}^2(d_{14},d_{24})
\eeq
The Grassmannian manifold is defined by the following embedding:
\beq
&& Gr_{4,2}^{(\underline{1,1},0,-1)}=\nn && \{W{\bf C}P^5_{(\underline{2,1,1},0,0,-1)}(d_{12},d_{13},d_{23},d_{14},d_{24},d_{34}) \ | \
d_{12}d_{34} - d_{13}d_{24} + d_{14}d_{23} = 0 \}. \label{homogembed}\eeq
 From the discussion in Appendix \ref{wpsnw} we know
that \nychanged{$W{\bf C}P^3_{(\underline{2,1,1},-1)}$ can be written}
as a blow-up along a plane with a $\mathbb{Z}_2$ action:
\beq W{\bf C}P^5_{(\underline{2,1,1},0,0,-1)}=\tilde{{\bf C}}^3/\mathbb{Z}_2
\times {\bf C}^2 \equiv \tilde{{\bf C}}_{x_4,x_5}^5 /\mathbb{Z}_2,
\eeq
where the right-most side \nychanged{means} that we blow-up
\nychanged{${\bf C}^5(x_1,x_2,x_3,x_4,x_5)$} along the plane
$x_1=x_2=x_3=0$, and the
$\mathbb{Z}_2$ action is given by $x_1=-x_1$. Far from the blown-up plane there is a one-to-one correspondence between the
homogeneous coordinate $d_{i j}$ and the inhomogeneous coordinate $x_i$:
\beq
\label{inhomogeneouscoord} x_1^2=d_{12} d_{34}^2,\quad x_2=d_{13} d_{34},\quad x_3=d_{23} d_{34},\quad x_4=d_{14},\quad
x_5=d_{24},
\eeq
and the Pl\"ucker relation becomes in the coordinates of ${\bf C}^5$:
\beq
\label{surf} x^2_1-x_5 x_2+ x_4 x_3=0,
\eeq
which \nychanged{defines} a cone inside ${\bf C}^5$, and
\nychanged{defines} our Grassmannian manifold as
the following embedding:
\beq
&& {\cal M}_{k=2,N=2,\NF=3}= WGr_{4,2}^{(\underline{1,1},0,-1)}= \nn && \{\tilde{{\bf
C}}_{x_4,x_5}^5(x_1,x_2,x_3,x_4,x_5)/\mathbb{Z}_2 \ | \ x^2_1-x_5 x_2+ x_4 x_3=0 \}. \label{blowupembedding}
\eeq
This enables us to consider $WGr_{4,2}^{(\underline{1,1},0,-1)}$ as the blow-up along a plane of the cone $x^2_1-x_5 x_2+ x_4
x_3=0$.

Consider now a local vortex. This case corresponds to the vanishing of the three minors:
\beq
d_{14}=d_{24}=d_{34}=0
\eeq
The embedding relation is trivially satisfied, and from:
\beq
W{\bf C}P^3_{(\underline{2,1,1},-1)}(d_{12},d_{13},d_{23},0) \times {\bf C}^2(0,0)=W{\bf
C}P^2_{(2,1,1)}(d_{12},d_{13},d_{23})={\bf C}P^2/\mathbb{Z}_2
\eeq
 we recover the correct answer for the moduli space of local vortices. In the language of \nychanged{(\ref{blowupembedding})} the condition for
local vortices implies $x_1=x_2=x_3=x_4=x_5=0$, e.g the origin of $\tilde{{\bf C}}^5$. This point must be blown-up, and it follows that
it is mapped into a ${\bf C}P^2/\mathbb{Z}_2=W{\bf C}P^2_{(2,1,1)}$. In the semilocal case we have $x_3,x_4,x_5\neq
0$. The surface \nychanged{(\ref{surf})} now is nontrivial, and passes trough the plane $x_4,x_5= 0$, that is blown-up.

\item \underline{$\NC=1, \NF=3$}

In this case we collect our matrices into the following one:
\beq
M=\left(
  \begin{array}{cccc}
    1 & 0 & q_1 & q_2 \\
    0 & 1 & -p_1 & -p_2 \\
  \end{array}
\right)
\eeq
\end{itemize}
This can be thought as the only patch of the "regularized" weighted Grassmannian $WGr_{4,2}^{(1,1,0,\underline{-1})}$. This space
is simply:
\beq
{\cal M}_{k=2,N=1,\NF=3}={\bf C}^4(q_1,p_1,q_2,p_2). \label{abelian1}\eeq

Note that if we exchange the role of ${\bf \Psi}$ and $\tilde {\bf \Psi}$ we get a fourth patch for the Grassmannian, and we can
complete (\ref{grassmcollect}):
\beq
&\left(
\begin{array}{c}
d_{12}\\
d_{23}\\
d_{13}\\
d_{14}\\
d_{24}\\
d_{34}
\end{array}
\right) \equiv \left(
\begin{array}{c}
-a\\
1\\
b\\
-b q - ap\\
-q\\
-p
\end{array}
\right) \sim \left(
\begin{array}{c}
1\\
Y\\
X\\
-\eta\\
-\xi\\
Y\eta - X\xi
\end{array}
\right)\nn & \sim \left(
\begin{array}{c}
a'\\
b'\\
1\\
-q'\\
-b'q' - a'p'\\
-p'
\end{array}
\right)\sim \left(
\begin{array}{c}
- q_1 p_2+q_2 p_1\\
p_2\\
p_1\\
q_1\\
q_2\\
1
\end{array}
\right). \label{grassmcollectcomplete}
\eeq

\subsection{$ WGr_{4,2}^{(1,1,0,-1)}$ and duality}
In this section we give a description of the full, non-Hausdorff, $ WGr_{4,2}^{(1,1,0,-1)}$. This can be easily done solving the
constraint \nychanged{(\ref{surf})} for $x_1$. This is possible thanks to the $\mathbb{Z}_2$ action $x_1=-x_1$, so that the good coordinate is
just $x_1^2$:
\beq
\label{surf1} x^2_1=x_5 x_2- x_4 x_3.
\eeq
Thus, when $(x_2,x_3) \neq 0$ our moduli space is isomorphic to
${\bf C}^4 (x_2,x_3, x_4, x_5)$. When $(x_2,x_3)=0$, from
\nychanged{(\ref{surf1})} we get $x_1=0$ and this implies
$d_{34}=0$. This can be seen from \nychanged{(\ref{inhomogeneouscoord})}
and remembering that our
definition of $Gr_{4,2}^{(\underline{1,1},0,-1)}$ is such that $d_{12},d_{13},d_{23} \neq 0$. When $d_{34}=0$, \nychanged{the original definition (\ref{homogembed})
reduces} to:
\beq
\{ W{\bf C}P^2_{(2,1,1)}(d_{12},d_{13},d_{23},0) \times {\bf C}^2(x_4,x_5) \ | \ d_{13} x_5 = x_4 d_{23} . \}
\label{homogembedred}\eeq We see that the action of the blow-up
\nychanged{(\ref{blowupembedding})} on ${\bf C}^4 (x_2,x_3, x_4, x_5)$ is to
substitute the plane $x_2=x_3=0 \ (d_{34}=0)$ with the space
\nychanged{(\ref{homogembedred})}. In other words:
\nychanged{
\beq
&&{\cal M}_{k=2,N=2,\NF=3}=WGr_{4,2}^{(\underline{1,1},0,-1)}={\bf C}^{* 2}(x_2,x_3) \times {\bf C}^2(x_4,x_5) \nonumber \\
&\oplus&\{ W{\bf C}P^2_{(2,1,1)}(d_{12},d_{13},d_{23}) \times {\bf C}^2(x_4,x_5) \ | \ d_{13} x_5 = x_4 d_{23}~(x_2=x_3=d_{34}= 0) \}. \label{oplusmoduli}
\eeq
}
In the lump limit we take $p \neq 0$, that is $d_{34} \neq 0$. From \ref{oplusmoduli}:
\beq
{\cal M}_{k=2,N=2,\tNC=1}^{{\rm lump}}=WGr_{4,2}^{(\underline{1,1},0,\underline{-1})}={\bf C}^{* 2}(x_2,x_3) \times {\bf C}^2(x_4,x_5)
\eeq

We can rewrite \nychanged{(\ref{oplusmoduli})} as the set of points inside:
\beq \label{moduliNc=2}
W{\bf C}P^3_{(\underline{2,1,1},-1)}(d_{12},d_{13},d_{23},d_{34})\times {\bf C}^2(x_2,x_3) \times {\bf C}^2(x_4,x_5)
\eeq
that satisfy the relations:
\beq
d_{13}d_{34}=x_2, \quad d_{23} d_{34}=x_3, \quad d_{12} d_{34}-d_{13}x_5+d_{23} x_4=0
\eeq

The moduli space for the dual theory can be obtained from the previous expression just substituting $W{\bf
C}P^3_{(\underline{2,1,1},-1)}$ with $W{\bf C}P^3_{(2,1,1,\underline{-1})}$. But now the condition $x_2=x_3=0$ does not implies
$d_{34}=0$, but instead $d_{12}=d_{13}=d_{23}=0$, thus:
\nychanged{
\beq
 {\cal M}_{k=2,N=1,\NF=3}=WGr_{4,2}^{(1,1,0,\underline{-1})}={\bf C}^{* 2}(x_2,x_3) \times {\bf C}^2(x_4,x_5) \nonumber\\
\oplus~~\{ W{\bf C}P^2_{(2,1,1,\underline{-1})}(0,0,0,d_{34}) \times {\bf C}^2(x_4,x_5)~(x_2=x_3= 0) \}.
\label{oplusmoduli2}
\eeq
}
That is, of course:
\beq
 {\cal M}_{k=2,N=1,\NF=3}={\bf C}^{* 2}(x_2,x_3) \times {\bf C}^2(x_4,x_5) \oplus (0,0) \times {\bf C}^2(x_4,x_5)= {\bf
C}^4(x_1,x_2,x_4,x_5).
 \label{abelian2}\eeq Note that (\ref{inhomogeneouscoord}) give, in the patch where we put $d_{34}=1$ (thanks to
\ref{grassmcollectcomplete}) :
\beq
x_2=d_{13}=p_1, \quad x_3=d_{23}=p_2, \quad x_4=d_{14}=q_1, \quad x_5=d_{24}=q_2.
\eeq
So that (\ref{abelian}) and (\ref{abelian2}) are consistent.

Note that we can also consistently use transition functions from (\ref{grassmcollectcomplete}) to identify non-normilizable modes
given those of the dual system.

\subsection{General case}

The general case with $\tNC$ flavors involves weighted spaces with several negative weights:

\beq WGr_{\NC+\tNC+1,2}^{(1_{\NC},0,-1_{\NC})} \subset W{\bf
C}P^{I+\NF-1}_{(2_I,1_{\NC},-1_{\tNC})}\times {\bf C}^{\NC \tNC},\quad I=\NC(\NC-1)/2.
\eeq

This is Calabi-Yau for $\NC^2=\tNC$.

\section{Moduli space of lump in terms of moduli matrix}\label{Ohashi}

In this appendix we will show that the moduli space of lumps is given by:
\begin{eqnarray}
{\cal M}_{\NC,\NF;k}^{{\rm lump}}= \left\{ ({\bf Z,\Psi,\widetilde \Psi})\big| \; GL(k,{\bf C})\; {\rm free\; on}\; ({\bf Z,\Psi})\;
{\rm and}\; ({\bf Z,\widetilde \Psi}) \right\}/GL(k,{\bf C}).
\end{eqnarray}
Recall that we start from the situation where $GL(k,{\bf C})$ acts freely on $({\bf Z,\Psi})$ only. This means that we have to prove the following:

{\bf Theorem.} The following two statements are equivalent:
\begin{itemize}
  \item[i)] $({\bf Z,\widetilde \Psi}):{GL(k,{\bf C})\rm ~free~}$
  \item[ii)] ${}^\forall z:\quad \det H_0(z)H_0(z)^\dagger\not =0$.
\end{itemize}

Let us begin with some preliminary considerations:

 \emph{Lump condition}. Let us decompose the moduli matrix like in Eq.~(\ref{decomp}):
\beq H_0(z)=\left({\bf D}(z),{\bf Q}(z)\right )
\eeq
\kochanged{The rational map (\ref{rationalmap}) gives,
\begin{eqnarray}
{\bf D}(z)^{-1}H_0(z)=\left({\bf 1}_{N_{\rm C}},\,{\bf R}(z)\right)=
\left({\bf 1}_{N_{\rm C}},\,P(z)^{-1}{\bf F}(z)\right) =\left({\bf 1}_{N_{\rm C}},\, {\bf
\Psi}\frac{1}{z-{\bf Z}}\widetilde {\bf \Psi}\right) . \label{eq:DH}
\end{eqnarray}
}
The lump condition is thus equivalent to:
\begin{eqnarray}
\quad {}^\forall z:\quad \det H_0(z)H_0(z)^\dagger=|P(z)|^2\det\left({\bf 1}_{N_{\rm C}}+|P(z)|^{-2}{\bf F}(z){\bf
F}^{\dagger}(z)\right)\neq0,\label{h0condition}
\end{eqnarray}
When $z\neq z_i$, $P(z)\neq0$ and the argument of the determinant is positive definite. The condition above is thus equivalent
to the following:
\begin{eqnarray}
{}^\forall i: \quad |P(z)|^2\det\left({\bf 1}_{N_{\rm C}}+|P(z)|^{-2}{\bf F}(z){\bf F}^{\dagger}(z)\right)\nrightarrow 0 \quad
{\rm for } \, z \rightarrow z_i.
\end{eqnarray}

We can give another more convenient form for the lump condition. Let us consider the following matrices (here we use the
notations defined in the next subsection):

\begin{eqnarray}
F^{\tilde r_1\tilde r_2\cdots \tilde r_I}_{s_1s_2\cdots s_I}(z)& \equiv&
\sum_{\{r_1,r_2,\cdots,r_{N_{\rm C}-I}\}}\frac1{(N_{\rm C}-I)!} \epsilon_{s_1s_2\cdots
s_Ir_1r_2\cdots r_{N_{\rm C}-I}} \det H_0^{\langle r_1r_2\cdots r_{N_{\rm C}-I} \tilde r_1\tilde r_2\cdots \tilde r_I
\rangle}(z)\nn & =&
\sum_{\{r_i\}}\frac{P(z)}{(N_{\rm C}-I)!} \epsilon_{s_1s_2\cdots s_Ir_1r_2\cdots r_{N_{\rm C}-I}} \det
(D(z)^{-1}H_0(z))^{\langle r_1r_2\cdots r_{N_{\rm C}-I} \tilde r_1\tilde r_2\cdots \tilde r_I \rangle} \nn & =&P(z) \, \det
\left({\bf \Psi}\frac{1}{z-{\bf Z}}\widetilde {\bf \Psi} \right)_{\langle \{s\}\rangle}{}^{\langle \{\tilde r\} \rangle}\nn &
=&P(z)\det{\bf \Psi}_{\langle \{s\}\rangle}{}^{\langle \{i\}\rangle} \det\left(\frac{1}{z-{\bf Z}}\right)_{\langle \{i\}\rangle }
{}^{\langle \{j\}\rangle} \det \widetilde {\bf \Psi}_{\langle \{j\}\rangle}{} ^{\langle \{\tilde r\} \rangle}.\label{fdef}
\end{eqnarray}

With $s_i,r_i\in \{1,2,\cdots ,N_{\rm C}\}$, $\tilde r_i\in \{N_{\rm C}+1,\cdots, N_{\rm F} \}$ and $1\le I\le {\rm min}(N_{\rm
C},\, \tilde N_{\rm C},\,k)$. The matrices $F^{\tilde r_1\tilde r_2\cdots \tilde r_I}_{s_1s_2\cdots s_I}(z)$ are a kind of
generalization of the matrix ${\bf F}(z)$. In fact we have: $F^{\tilde r_1}_{s_1}(z)={\bf F}(z)$. In fact the only independent
quantities are that with $I=0$, ${\bf P}(z)$, and $I=1$, ${\bf F}(z)$. The other matrices are related to the former ones by
homogeneous relations (Plucker conditions). Now, using the identity (\ref{detidentity}) we can translate the condition
(\ref{h0condition}) into the following:
\begin{eqnarray}
\quad {}^\forall z,\quad {}^\exists \{A\} :\quad \det H_0^{\langle\{A\}\rangle}(z)\not =0,
\end{eqnarray}
that is equivalent to:
\begin{eqnarray}
{}^\forall i, \quad{}^\exists \{\tilde r\},\,{}^\exists \{s\} :\quad F^{\{\tilde r\}}_{\{s\}}(z_i)\not=0.
\end{eqnarray}
We can rephrase our theorem in the most convenient form.

{\bf Theorem.} The following two statements are equivalent:
\begin{itemize}
  \item[i)] $({\bf Z,\widetilde \Psi}):{GL(k,{\bf C})\rm ~free~}$
  \item[ii)] ${}^\forall i, \quad{}^\exists \{\tilde r\},\,{}^\exists \{s\} :\quad F^{\{\tilde r\}}_{\{s\}}(z_i)\not=0$.
\end{itemize}

\emph{Jordan form and $GL(k,{\bf C})$ free condition}. ${\bf Z}$ can always be set to a canonical block-wise form \`a la Jordan,
that is in our choice lower-triangular:
\begin{eqnarray}
{\bf Z}=\left(
\begin{array}{cccc}
{\bf Z}_1& {\bf 0}&\cdots & \\
{\bf 0}& {\bf Z}_2&&\\
\vdots &&\ddots\\
&&&{\bf Z}_N
\end{array}\right), \quad
{\bf Z}_i=\left(
\begin{array}{cccc}
z_i& 0&\cdots & \\
1& z_i&&\\
0 &\ddots&\ddots\\
\vdots&&1& z_i
\end{array}\right),\label{eigendecomp}
\end{eqnarray}
where ${\bf Z}_i\,(i=1,\cdots,N)$ is an $\alpha_i\times \alpha_i$ block and $\sum_{i=1}^N\alpha_i=k$. Here two eigenvalues $z_i,
z_j$ are allowed to be the same for $i\neq j$. It is convenient also to
define sets of indices, $I_a\, (a=1,\cdots,d)$,
that collect ${\bf Z}_i$ blocks with the same eigenvalue:
\begin{eqnarray}
 \bigcup_{a=1}^d I_a=\{1,\cdots,N\},\quad
I_a\cap I_b=\emptyset \quad {\rm for~} a\not=b;\nn
{}^\forall i,j\in I_a,\quad z_i=z_j
\equiv z_{(a)};\nn {}^\forall i\in I_a, {}^\forall j\in I_b \, (b\not =a),\quad z_i\not =z_j.
\end{eqnarray}

We also decompose $\widetilde {\bf \Psi}$ into $\alpha_i\times \tilde N_{\rm C} $ matrix $\widetilde {\bf \Psi}_i$, from which we
extract the $|I_a|\times \tilde N_{\rm C}$ matrices $\widetilde {\bf A}_{(a)}$:
\begin{eqnarray}
\widetilde {\bf \Psi}=\left(
\begin{array}{c}
\widetilde {\bf \Psi}_1 \\ \widetilde {\bf \Psi}_2 \\
\vdots\\ \widetilde {\bf \Psi}_N
\end{array}\right),\quad
(\widetilde {\bf A}_{(a)})_n{}^{\tilde r}\equiv (\widetilde {\bf \Psi}_{i_n})_1{}^{\tilde r}, I_a=\{i_1,i_2,\cdots,i_{|I_a|}\}.
\label{atilde}\end{eqnarray} The matrices $(\widetilde {\bf A}_{(a)})_n{}^{\tilde r}$ simply collect the $|I_a|$ first rows of
the blocks $\widetilde {\bf \Psi}_{i}$ that corresponds to the same eigenvalue in the decomposition (\ref{eigendecomp}). With
these definitions we are ready to prove the following:

 {\bf Proposition 1.}~The following two statements are equivalent:
 \begin{itemize}
   \item[i)] $({\bf Z,\widetilde \Psi}):\quad {GL(k,{\bf C})\rm ~free~}
\label{eq:statement_1}$
   \item[ii)] $ {}^\forall a,{}^{\exists} \{\tilde r_a\}\subset\{N_{\rm C}+1,\cdots,N_{\rm F}\},\quad |\{\tilde r_a\}|=|I_a|:\quad \quad\det (\widetilde {\bf A}_{(a)})^{\langle \{\tilde r_a\} \rangle}\not=0. \label{eq:statement_2}$
 \end{itemize}

{\bf Proof 1.}~Let us consider a matrix ${\bf X}\in GL(k,{\bf C})$ satisfying:
\begin{eqnarray}
[{\bf Z},\,{\bf X}]=0, \quad {\bf X}\widetilde {\bf \Psi}=0, \label{eq:X}
\end{eqnarray}
and decompose it to smaller matrices as
\begin{eqnarray}
{\bf X}=\left(
\begin{array}{cccc}
{\bf X}_{11}& {\bf X}_{12}&\cdots & {\bf X}_{1N}\\
{\bf X}_{21}& \ddots&&\\
\vdots & &\ddots\\
{\bf X}_{N1}&&\cdots& {\bf X}_{NN}
\end{array}\right)
\end{eqnarray}
where ${\bf X}_{ij}$ is $\alpha_i\times \alpha_j$.

The first condition of Eq.~(\ref{eq:X}) is thus decomposed as:
\begin{eqnarray}
[{\bf Z},{\bf X}]=0\, \Leftrightarrow\, {\bf Z}_i{\bf X}_{ij}={\bf X}_{ij}{\bf Z}_j,
\end{eqnarray}
where the indices $i$ and $j$ are not summed over. Using the explicit form (\ref{eigendecomp}) for ${\bf Z}_i$, the above
condition gives the following recurrence formula:
\begin{eqnarray}
(z_i-z_j)({\bf X}_{ij})_p{}^q=({\bf X}_{ij})_p{}^{q+1} -({\bf X}_{ij})_{p-1}{}^{q},\label{recursion}
\end{eqnarray}
where $p=1,\cdots,\alpha_i$ and $q=1,\cdots,\alpha_j$, and we have defined $({\bf X}_{ij})_0{}^q\equiv0$ and $({\bf
X}_{ij})_p{}^{\alpha_j+1}\equiv0$. Especially this gives $(z_i-z_j)({\bf X}_{ij})_1{}^{\alpha_j}=0$. By use of
Eq.~(\ref{recursion}), we inductively find:
\begin{eqnarray}
{\bf X}_{ij}={\bf 0} \quad & &{\rm for~} z_i\not=z_j;\nn
 ({\bf X}_{ij})_p{}^q=x_{ij}^{(\alpha_i-p+q)} \quad & &{\rm for~}
z_i=z_j, \nn {\rm with }\quad x_{ij}^{(l)}=0\quad & &{\rm for~} l>\alpha_{ij}\equiv {\rm min}(\alpha_i,\alpha_j).
\end{eqnarray}
The $x_{ij}^{(l)}$ are the undetermined entries of the lower-triangular part of ${\bf X}_{ij}$, for example:
\beq
{\bf X}_{ij}=\left(
  \begin{array}{ccccccc}
    x_{ij}^{(\alpha_{ij})} & 0 & \cdots & \cdots & \cdots & \cdots & 0 \\
    \vdots & \ddots & \ddots & \cdots & \cdots & \cdots & 0 \\
    x_{ij}^2 & \ddots &\ddots & \ddots & \cdots & \cdots & 0 \\
    x_{ij}^1 & x_{ij}^2 & \cdots & x_{ij}^{(\alpha_{ij})} & 0 & \cdots & 0 \\
  \end{array}
\right).
\eeq
The second condition gives:
\begin{eqnarray}
{\bf X}\widetilde {\bf \Psi}=0 \quad \Rightarrow \quad &&\sum_{\{j|z_i=z_j\}}x_{ij}^{(\alpha_{ij})} (\widetilde {\bf
\Psi}_j)_1{}^{\tilde r}=0, \nn
 &&\sum_{\{j|z_i=z_j\}}x_{ij}^{(\alpha_{ij}-1)}(\widetilde {\bf \Psi}_j)_1{}^{\tilde r}
+x_{ij}^{(\alpha_{ij})}(\widetilde {\bf \Psi}_j)_2{}^{\tilde r}=0, \nn
 &&\sum_{\{j|z_i=z_j\}}x_{ij}^{(\alpha_{ij}-2)}(\widetilde
{\bf \Psi}_j)_1{}^{\tilde r} +x_{ij}^{(\alpha_{ij}-1)}(\widetilde {\bf \Psi}_j)_2{}^{\tilde r}+x_{ij}^{(\alpha_{ij})}(\widetilde
{\bf \Psi}_j)_3{}^{\tilde r}=0, \nn
 && \cdots
\end{eqnarray}

Now let us assume the statement ii) of Proposition 1. This implies that we have an invertible square sub-matrix $(\widetilde {\bf
A})_j{}^{\langle \{\tilde r\} \rangle}=(\widetilde {\bf \Psi}_j)_1{}^{\langle \{\tilde r\} \rangle}$. If we multiply the first
equation with the inverse of this matrix we get $x_{ij}^{(\alpha_{ij})}=0$. Plugging this result in the second equation we get
$x_{ij}^{(\alpha_{ij}-1)}=0$, and by induction we conclude that all $x_{ij}^{(l)}$ vanish and that ${\bf X}\equiv 0$. This
implies the statement i).

\kochanged{ Conversely, let us assume the statement ii) is false. This implies that there exists a non-vanishing matrix $y_{ij}$ satisfying $y\cdot
\widetilde {\bf A}=0$ and that we can find non-vanishing ${\bf X}$ that satisfies Eq.~(\ref{eq:X}). For instance we can
take: $x_{ij}^{(1)}=y_{ij}$ and the others vanish:
$x_{ij}^{(p)}=0\,(p>1)$.
Thus the statement i) is false. }

These arguments prove the proposition.
{\tiny $\blacksquare$}

Similarly we can prove:

{\bf Proposition 2.}~The following two statements are equivalent:
\begin{itemize}
  \item[i)] $({\bf Z,\Psi}):\quad {GL(k,{\bf C})\rm ~free~}
\label{eq:statement_3}$
  \item[ii)] ${}^\forall a,{}^{\exists} \{r_a\}\subset
\{1,\cdots,N_{\rm C}\},\quad |\{r_a\}|=|I_a|:\quad\det ({\bf A}_{(a)})_{\langle \{r_a\} \rangle}\not=0. \label{eq:statement_4}$
\end{itemize}

with the following decomposition and definition:
\begin{eqnarray}
{\bf \Psi}=\left( {\bf \Psi}_1\, {\bf \Psi}_2\,\cdots\, {\bf \Psi}_N\right),\quad ({\bf A}_{(a)})_{r}{}^{n}\equiv ({\bf
\Psi}_{i_n})_r{}^{\alpha_{i_n}}, I_a=\{i_1,i_2,\cdots,i_{|I_a|}\}.\label{a}
\end{eqnarray}

{\bf Proof 2.}~ The proof proceeds analogously to that of Proposition 1. Alternatively one can note that there exists $U\in
GL(k,{\bf C})$ satisfying $U{\bf Z}U^{-1}={\bf Z}^{\rm T}$. {\tiny $\blacksquare$}

\kochanged{Proposition 1 \nychanged{states} that
in regions with $|I_a|>\widetilde N_{\rm C}$, $GL(k,{\bf C})$ is always
non-free on $({\bf Z},\,{\bf \widetilde \Psi})$, since we cannot
choose an $|I_a|\times |I_a|$ matrix from the matrix $\bf A_{(a)}$.
Similary, according to Proposition 2, $|I_a|$ is always smaller than
$N_{\rm C}$ so that $GL(k,{\bf C})$ acts freely on $({\bf Z},\,{\bf \Psi})$.
}

{\bf Proof of the theorem}.

 For $z \ne z_i$ we can consider the matrix:
\begin{eqnarray}
\frac{{ 1}}{z-{\bf Z}}&=&
\left(
\begin{array}{ccc}
\frac{1}{z-{\bf Z_1}}& \\
&\frac{1}{z-{\bf Z_2}}\\
&&\ddots \\
\end{array}\right).
\end{eqnarray}
Note that ${\bf Z}_i=z_i{\bf 1}_{\alpha_i}+E_{(\alpha_i)}$,
with
\beq
E_{(\alpha_i)}=\left(
\begin{array}{cccc}
0& 0&\cdots & \\
1& 0&&\\
0 &\ddots&\ddots\\
\vdots&&1& 0
\end{array}\right),
\eeqq
that is the $\alpha_i$-nilpotent Jordan block, $E^{\alpha_i}_{(\alpha_i)}=0$. Thus we obtain
\begin{eqnarray}
\frac{1}{z-{\bf Z}_i}&=& \frac{1}{(z-z_i){\bf 1}_{\alpha_i}-E_{(\alpha_i)}}
=\frac{1}{z-z_i}\sum_{n=0}^{\alpha_i-1}\frac{E_{(\alpha_i)}^n}{(z-z_i)^n}=\nn &=&\left(
\begin{array}{cccc}
(z-z_i)^{-1}&0 &\cdots & 0\\
(z-z_i)^{-2}&(z-z_i)^{-1}&&\vdots\\
\vdots&\ddots&\ddots&0\\
(z-z_i)^{-\alpha_i}&\cdots&(z-z_i)^{-2}&(z-z_i)^{-1}
\end{array}\right).
\end{eqnarray}
It is easy to see that the minor determinants of this matrix either vanish identically or behave, in the vicinity of $z_i$, like:
\begin{eqnarray}
\det\left(\frac{1}{z-{\bf Z}_i}\right)_{\langle\{p\}\rangle} {}^{\langle\{q\}\rangle}\propto \frac1{(z-z_i)^s},\quad s\le
\alpha_i,
\end{eqnarray}
and the equality $s=\alpha_i$ implies $\alpha_i\in\{p\}$ and $1\in \{q\}$. The minimal size of a minor matrix whose determinant
goes like $(z-z_i)^{-\alpha_i}$ is $1\times 1$, and it is achieved by
$\{p\}=\{\alpha_i\}$ and $\{q\}=\{1\}$.
\kochanged{Therefore $F_{\{s\}}^{\{\tilde r\}}(z_{(a)})$ vanishes in the
case of $I<|I_{(a)}|$ and, in the case of
$I=|I_a|$, \nychanged{the} non-vanishing minor determinant, \nychanged{whose} size is $|I_a|$, is
given by,
\begin{eqnarray}
P(z)\det\left(\frac{1}{z-{\bf Z}}\right)_{\langle\{\nu^{(a)}\}\rangle} {}^{\langle\{\mu^{(a)}\}\rangle}\Big|_{z\rightarrow z_{(a)}}
=\prod_{b\not=a}(z_{(a)}-z_{(b)})^{n_{(b)}}\not=0,
\end{eqnarray}
where ${n_{(b)}}$ is the algebraic multiplicity of $z_{(b)}$ and, most
importantly, sets of $|I_a|$ indices
$\{\mu^{(a)}\}=\{\mu^{(a)}_1,\mu^{(a)}_2,\cdots\}$ and
$\{\nu^{(a)}\}=\{\nu^{(a)}_1,\nu^{(a)}_2,\cdots\}$
are chosen so that
${\bf \Psi}_s{}^{\nu_i^{(a)}}=({\bf A}_{(a)})_s{}^i$ and
${\bf \widetilde \Psi}_{\mu_j^{(a)}}{}^{\tilde r}=
({\bf \widetilde A}_{(a)})_j{}^{\tilde r}$.
Using \nychanged{(\ref{fdef}), (\ref{a})
and (\ref{atilde})} we get, for $|\{ \tilde r\}|=|\{s\}|=|I_a|$:
\begin{eqnarray}
F^{\{\tilde r\}}_{\{s\}}(z_{(a)})=\prod_{b\not=a}(z_{(a)}-z_{(b)})^{n_{(b)}} \det({\bf A}_{(a)})_{\langle\{s\}\rangle}
\det(\widetilde{\bf A}_{(a)})^{\langle\{\tilde r\}\rangle}.\label{fidentity}
\end{eqnarray}
In case of $I>|I_a|$, $\alpha_i \in {p}$ and $1 \in {q}$ implies that
the minor determinant always choose all lines of the matrix ${\bf
A}_{a}$, that is,
\begin{eqnarray}
 F^{\{\tilde r\}}_{\{s\}}(z_{(a)})&=&\sum_{J\supset \{\mu^{(a)}\},|J|=I}
C_{\langle\{s\}\rangle}{}^{\langle J\rangle}
\det \widetilde {\bf \Psi}_{\langle J\rangle}{}^{\langle \{\tilde r\}
\rangle}\nn
&=&\sum_{J}\sum_{K\subset\{\tilde r\}, |K|=|I_a|}
C'_{\langle\{s\}\rangle}{}^{\langle J\rangle}
\det(\widetilde{\bf A}_{(a)})^{\langle K\rangle}
\det \widetilde {\bf \Psi}_{\langle J-\{\mu^{(a)}\}\rangle}
{}^{\langle \{\tilde r\} -K\rangle}
\end{eqnarray}
where $C_{\langle\{s\}\rangle}{}^{\langle\{\mu\}\rangle}$ and
$C'_{\langle\{s\}\rangle}{}^{\langle\{\mu\}\rangle}$ are certain
constants.
}

\kochanged{Thus we find that the existence of a non-vanishing
$\det \widetilde {\bf A}_{(a)}$ means that there exists an non-vanishing
$F(z_{(a)})$.
 And if all of $\det \tilde {\bf A}_{(a)}$ vanish, then all
 $F(z_{(a)})$ vanish. According to Proposition 1,
these facts give a proof of the theorem. {\tiny $\blacksquare$}
}

\subsection{\kochanged{Minor Determinants}}\label{sec:identities}
Let us define the following notation for a determinant of a minor matrix:
\begin{eqnarray}
{\det } X^{\langle a_1a_2\cdots a_n\rangle} {}_{\langle b_1b_2\cdots b_n\rangle}\equiv n! X^{[a_1}{}_{b_1}X^{a_2}{}_{b_2}\cdots
X^{a_n]}{}_{b_n},
\end{eqnarray}
where $X$ is an $N\times M$ matrix and $n\le {\rm min}(M,N)$. Note that if $N=M$ we get the usual expression for the
determinant of a square matrix:
\begin{eqnarray}
{\det } X={\det } X^{\langle 1 2\cdots N\rangle} {}_{\langle 12\cdots N\rangle}.
\end{eqnarray}
We also use the following abbreviations:
\begin{eqnarray}
{\det } X^{\langle a_1a_2\cdots a_N\rangle} &=&{\det } X^{\langle 1 2\cdots N\rangle}{}_{\langle a_1a_2\cdots a_N\rangle}, \quad
{\rm for~}{N<M};\nn {\det } X_{\langle a_1a_2\cdots a_M\rangle}&=& {\det } X^{\langle a_1a_2\cdots a_M\rangle}{}_{\langle 1
2\cdots M\rangle}, \quad {\rm for~}{N>M}.
\end{eqnarray}
We used the following useful identity:
\begin{eqnarray}
{\det }(XY)^{\langle a_1a_2\cdots a_n\rangle} {}_{\langle b_1b_2\cdots b_n\rangle}&=& n!
(XY)^{[a_1}{}_{b_1}(XY)^{a_2}{}_{b_2}\cdots (XY)^{a_n]}{}_{b_n}\nn &=&n! X^{[a_1}{}_{c_1}X^{a_2}{}_{c_2}\cdots X^{a_n]}{}_{c_n}
Y^{c_1}{}_{b_1}Y^{c_2}{}_{b_2}\cdots Y^{c_n}{}_{b_n}\nn &=&\frac1 {n!}{\det } X^{\langle a_1a_2\cdots a_n\rangle} {}_{\langle
c_1c_2\cdots c_n\rangle}{\det } Y^{\langle c_1c_2\cdots c_n\rangle} {}_{\langle b_1b_2\cdots b_n\rangle},
\end{eqnarray}
\kochanged{where Einstein contraction
with respect to a set of indices $\{c_1,c_2,\cdots\}$ has been assumed.
This equation can be rewritten shortly \nychanged{as}
}
\begin{eqnarray}
{\det }(XY)^{\langle \{a\}\rangle} {}_{\langle \{b\}\rangle}={\det }X^{\langle \{a\}\rangle} {}_{\langle \{c\}\rangle}{\det
}Y^{\langle \{c\}\rangle} {}_{\langle \{b\}\rangle}
\end{eqnarray}
where $\langle \{a\}\rangle\equiv \langle a_1a_2\cdots a_n\rangle$ with $a_1<a_2<\cdots<a_n$. For instance, with $N<M$
\begin{eqnarray}
\det XX^\dagger= \det X_{\langle \{a\}\rangle}\det (X^\dagger)^{\langle \{a\}\rangle} =\sum_{\langle \{a\}\rangle}\big|\det
X_{\langle \{a\}\rangle}\big|^2\label{detidentity}
\end{eqnarray}

\end{document}